\documentclass[12pt]{article}

\advance\voffset by -2.0cm \advance\hoffset by -1.25cm
\usepackage{amsmath}
\usepackage{amssymb}
\usepackage{amsthm}

\newcommand{\CC}{\mathbb{C}} 
\newcommand{\RR}{\mathbb{R}} 
 
\newcommand{\FF}{\mathbb{F}} 
\newcommand{\ZZ}{\mathbb{Z}}

\newcommand{\MM}{\mathbb{M}}
\newcommand{\Hh}{\mathcal{H}}

\newcommand{\N}{\mathcal{N}}

\newcommand{\TT}{\mathbb{T}}

\newcommand{\C}{\mathcal{C}}

\newcommand{\g}{\mathfrak{g}}
\newcommand{\IZ}{{\mathbb Z}}

\DeclareMathOperator{\Tr}{Tr}

\DeclareMathOperator{\rk}{rk}

\numberwithin{equation}{section}

\def\be{\begin{equation}}
\def\ee{\end{equation}}

\allowdisplaybreaks[3]

\author{Matthias R. Gaberdiel \\
Institut f\"ur Theoretische Physik \\
ETH Zurich, 8093 Zurich, Switzerland \\
\\
and \\
\\
Roberto Volpato\\
Albert-Einstein Institut f\"ur Gravitationsphysik \\
Am M\"uhlenberg 1, 14476 Golm, Germany}
\title{Mathieu Moonshine and Orbifold K3s\footnote{Partially based on 
talk given by M.R.G.\ at the conference 
`Conformal Field Theory, Automorphic Forms and Related Topics', 
Heidelberg, 19-23 September 2011.}}

\begin{document}

%\begin{titlepage}

\begin{flushright}
AEI-2012-058
\end{flushright}

\vspace{8mm}

\begin{center}
{\LARGE Mathieu Moonshine and Orbifold K3s}\footnote{Partially based on 
talk given by M.R.G.\ at the conference 
`Conformal Field Theory, Automorphic Forms and Related Topics', 
Heidelberg, 19-23 September 2011.}
%\end{center}

\vspace{8mm}

Matthias R. Gaberdiel \\
Institut f\"ur Theoretische Physik \\
ETH Zurich, 8093 Zurich, Switzerland \\
\bigskip
and

\bigskip
Roberto Volpato\\
Max-Planck-Institut f\"ur Gravitationsphysik \\
Am M\"uhlenberg 1, 14476 Golm, Germany
%\abstract{}
\end{center}

\vspace{1cm}

\noindent 
{\bf Abstract:} The current status of `Mathieu Moonshine', the idea that 
the Mathieu group $\mathbb{M}_{24}$ organises the  elliptic genus 
of K3, is reviewed. While there is a consistent
decomposition of all Fourier coefficients of the elliptic genus 
in terms of Mathieu $\mathbb{M}_{24}$ representations, a conceptual 
understanding of this phenomenon in terms of K3 sigma-models
is still missing. In particular, it follows from the recent classification of 
the automorphism groups of arbitrary K3 sigma-models that (i) 
there is no single K3 sigma-model that has $\mathbb{M}_{24}$
as an automorphism group; and (ii)  there exist `exceptional' K3 sigma-models whose 
automorphism group is not even a subgroup of $\mathbb{M}_{24}$.  
Here we show that all cyclic torus orbifolds are exceptional in this sense, and that 
almost all of the exceptional cases are  realised as cyclic torus orbifolds.
We also provide an explicit construction of a $\mathbb{Z}_5$ torus
orbifold that realises one exceptional class of K3 sigma-models.

\section{Introduction}

\setcounter{footnote}{1}

In 2010, Eguchi, Ooguri and Tachikawa observed that the elliptic genus of K3 shows
signs of an underlying Mathieu $\mathbb{M}_{24}$ group action \cite{EOT}. In 
particular, they noted (see section~\ref{sec:MatMoon} below for more details) that the Fourier coefficients of the 
elliptic genus can be written as sums of dimensions of irreducible
$\mathbb{M}_{24}$ representations.\footnote{Actually, they did not just look at the 
Fourier coefficients themselves, but at the decomposition of the elliptic genus
with respect to the elliptic genera of irreducible ${\cal N}=4$  superconformal
representations. They then noted that these expansion coefficients (and hence
in particular the usual Fourier coefficients) are sums of dimensions of irreducible
$\mathbb{M}_{24}$ representations.} This intriguing
observation is very reminiscent of the famous realisation of McKay and Thompson 
who noted that the Fourier expansion coefficients of the $J$-function can be 
written in terms of dimensions of representations of the Monster group
\cite{Thompson,CN}. This led to a 
development that is now usually referred to as `Monstrous Moonshine', see \cite{Gannon2006}  
for a nice review. One important upshot of that analysis was that the $J$-function can be thought 
of as the partition function of a self-dual conformal field theory, the `Monster conformal field 
theory' \cite{Borcherds,FLM}, whose automorphism group is precisely the Monster group.
The existence of this conformal field theory explains many aspects of Monstrous Moonshine
although not all --- in particular, the genus zero property is rather mysterious from
this point of view.

In the Mathieu case, the situation is somewhat different compared to the early days of
Monstrous Moonshine. It is by construction clear that the underlying conformal field theory 
{\em is} a K3 sigma-model (describing string propagation on the 
target space K3). However, this does 
not characterise the corresponding 
conformal field theory uniquely as there are many inequivalent such sigma-models --- in 
fact, there is an $80$-dimensional moduli space of such theories, all of which
lead to the same elliptic genus. The natural analogue
of the `Monster conformal field theory' would therefore be a special K3 sigma-model whose
automorphism group coincides with $\mathbb{M}_{24}$. Unfortunately, as we shall
review here (see section~\ref{sec:Theorem}), such a sigma-model does not exist: we have 
classified the 
automorphism groups of all K3 sigma-models, and none of them contains $\mathbb{M}_{24}$ 
\cite{Gaberdiel:2011fg}. In fact, not even all automorphism groups are contained 
in $\mathbb{M}_{24}$: the exceptional cases are the possibilities (ii), (iii) and (iv) of the 
theorem in section~\ref{sec:Theorem} (see \cite{Gaberdiel:2011fg}), as well as case (i) for 
nontrivial $G'$. Case (iii) was already shown in 
\cite{Gaberdiel:2011fg} to be realised by a specific Gepner model that is believed to 
be equivalent to a torus orbifold by $\mathbb{Z}_3$. 
%The other two examples considered in \cite{Gaberdiel:2011fg} were again torus orbifolds whose automorphism groups realised exceptional groups in case (i). 
Here we show that also cases
(ii) and (iv) are realised by actual K3s --- the argument in \cite{Gaberdiel:2011fg} 
for this relied on some assumption about the regularity of K3 sigma-models --- 
and in both cases the relevant K3s are again torus orbifolds. More specifically, 
case (ii) is realised by an asymmetric $\mathbb{Z}_5$ orbifold of 
$\mathbb{T}^4$ (see section~\ref{sec:Z5}),\footnote{Since the orbifold action is asymmetric, 
this evades various no-go-theorems (see e.g.\ \cite{Walton}) 
that state that the possible orbifold groups
are either $\mathbb{Z}_2$, $\mathbb{Z}_3$, $\mathbb{Z}_4$, or
$\mathbb{Z}_6$.}
 while for case (iii) the relevant orbifold is by $\mathbb{Z}_3$ (see section~\ref{sec:Z3}).

Cyclic torus orbifolds are rather special K3s since they always possess a 
quantum symmetry whose orbifold leads back to $\mathbb{T}^4$. Using
this property of cyclic torus orbifolds, we show (see section~\ref{sec:symme}) that the group of 
automorphisms of K3s that are cyclic torus orbifolds is always exceptional; in 
particular, the quantum symmetry itself is never an element of $\MM_{24}$. 
Although some `exceptional' automorphism groups (contained in case (i) of the
classification theorem) can also arise in K3 models 
that are not cyclic torus orbifolds, our observation may go a certain way towards 
explaining why only $\mathbb{M}_{24}$ seems to appear in the elliptic genus of K3. 
\smallskip

We should mention that Mathieu Moonshine can also be formulated in terms
of a mock modular form that can be naturally associated to the elliptic genus
of K3 \cite{EOT,Cheng:2010pq,CD1,CD2}; this point of view has recently led to an 
interesting class of generalisations \cite{CDH}. There are also indications that, just
as for Monstrous Moonshine, Mathieu Moonshine can possibly be understood in 
terms of an underlying Borcherds-Kac-Moody algebra 
\cite{Govindarajan:2010fu,Suresh,Govindarajan:2011em,Hohenegger:2011us}.

\section{Mathieu Moonshine}\label{sec:MatMoon}

Let us first review the basic idea of `Mathieu Moonshine'. We consider a conformal
field theory sigma-model with target space K3. This theory has ${\cal N}=(4,4)$
superconformal symmetry on the world-sheet. As a consequence, the space of 
states can be decomposed into representations of the ${\cal N}=4$ superconformal
algebra, both for the left- and the right-movers. (The left- and right-moving actions commute,
and thus we can find a simultaneous decomposition.) The full space of states 
takes then the form
\be\label{K3spec}
{\cal H} = \bigoplus_{i,j} N_{ij} \, {\cal H}_i \otimes \bar{\cal H}_j \ ,
\ee
where $i$ and $j$ label the different ${\cal N}=4$ superconformal representations,
and $N_{ij}\in{\mathbb N}_0$ denote the multiplicities with which these representations
appear. The ${\cal N}=4$ algebra contains, apart from the Virasoro algebra $L_n$ at
$c=6$, four supercharge generators, as well as an affine $\hat{\mathfrak{su}}(2)_1$ 
subalgebra at level one; we denote the Cartan generator of the zero mode subalgebra
$\mathfrak{su}(2)$ by $J_0$.

The full partition function of the conformal field theory is quite complicated, and is only
explicitly known at special points in the moduli space. However, there exists some sort
of partial index that is much better behaved. This is the so-called {\em elliptic genus} 
that is defined by
\be
\phi_{\rm K3}(\tau,z) = {\rm Tr}_{\rm RR} \Bigl(
q^{L_0 - \frac{c}{24}} \, y^{J_0} \, (-1)^F \, 
\bar{q}^{\bar{L}_0 - \frac{\bar{c}}{24}}\, (-1)^{\bar{F}} \Bigr)  \equiv \phi_{0,1}(\tau,z) \ .
\ee
Here the trace is only taken over the Ramond-Ramond part of the spectrum
(\ref{K3spec}), and the right-moving ${\cal N}=4$ modes are denoted by a bar. Furthermore,
$q=\exp(2\pi i \tau)$ and $y=\exp(2\pi i z)$, $F$ and $\bar{F}$ are the left- and right-moving fermion
number operators, and the two central charges equal
$c=\bar{c}=6$. Note that the elliptic genus
does not actually depend on $\bar\tau$, although $\bar{q}=\exp(-2\pi i \bar\tau)$ does; 
the reason for this is that, with respect to the right-moving algebra, the elliptic genus is like a
Witten index, and only the right-moving ground states contribute. To see this one notices that
states that are not annihilated by a supercharge zero mode appear always as a 
boson-fermion pair; the contribution of such a pair to the elliptic genus however vanishes
because the two states contribute with the opposite sign (as a consequence of the
 $(-1)^{\bar{F}}$ factor). Thus only the right-moving ground states, i.e.\ the states that are 
 annihilated  by all right-moving supercharge zero modes, contribute to the elliptic genus, 
 and the commutation relations of the ${\cal N}=4$ algebra then imply that 
they satisfy $(\bar{L}_0 -\tfrac{\bar{c}}{24}) \phi_{\rm ground}  = 0$; thus it follows that the elliptic 
genus is independent of $\bar{\tau}$. Note that this argument does not 
apply to the left-moving contributions because of the $y^{J_0}$ factor. (The supercharges
are `charged' with respect to the $J_0$ Cartan generator, and hence the two terms of 
a boson-fermion pair come with different powers of $y$. However, if we also set $y=1$,
the elliptic genus does indeed become a constant, independent of $\tau$ and 
$\bar\tau$.)

It follows from general string considerations that the elliptic genus defines a 
{\em weak Jacobi form of weight zero and index one} \cite{Kawai:1993jk}. Recall  
that a weak  Jacobi form of weight $w$ and index $m$ is a function \cite{EichlerZagier}
\be
\phi_{w,m} : {\mathbb H}_+ \times {\mathbb C} \rightarrow {\mathbb C}  \ , 
\qquad 
(\tau,z) \mapsto \phi_{w,m}(\tau,z) 
\ee
that satisfies
\begin{equation}\label{eq:jactmn1}
 \phi_{w,m}\Bigl(\frac{a \tau + b}{c \tau + d} , \frac{z}{c \tau + d}\Bigr) =
(c \tau+d)^w \, e^{ 2 \pi i m \frac{c z^2}{c \tau + d} } \, \phi_{w,m}(\tau,z)
\qquad \begin{pmatrix} a & b \\ c & d \\ \end{pmatrix} \in {\rm SL}(2,\IZ) \ ,
\end{equation}
\begin{equation}\label{eq:jactmn2}
 \phi(\tau,z+ \ell \tau + \ell') = e^{-2 \pi i m
(\ell^2 \tau+ 2 \ell z)} \phi(\tau,z) \qquad\qquad\qquad
\ell,\ell'\in \IZ \ ,
\end{equation}
and has a Fourier expansion
\begin{equation}
 \phi(\tau,z) = \sum_{n \geq 0,\, \ell\in \IZ} c(n,\ell) q^n
y^\ell \end{equation}
with $c(n,\ell) = (-1)^w c(n,-\ell)$. Weak Jacobi forms have been classified, and there is only 
one weak Jacobi form with $w=0$ and $m=1$. Up to normalisation $\phi_{\rm K3}$ must
therefore agree with this unique weak Jacobi form $\phi_{0,1}$, which can explicitly be
written in terms of Jacobi theta functions as 
\be
\phi_{0,1}(\tau,z) = 8 \sum_{i=2,3,4} \frac{\vartheta_i(\tau,z)^2}{\vartheta_i(\tau,0)^2} \ .
\ee
Note that the Fourier coefficients  of $\phi_{{\rm K3}}$ are integers; as a consequence
they cannot change continuously as one moves around in the moduli space of K3 
sigma-models, and thus $\phi_{{\rm K3}}$ must be actually independent of the 
specific K3 sigma-model  that is being considered, i.e.\ independent of the point in 
the moduli space. Here we have used that the moduli space is connected.
More concretely, it can be described as the double quotient
\be\label{moduli}
{\cal M}_{\rm K3} = {\rm O}(\Gamma^{4,20}) \  \backslash \  {\rm O}(4,20) \  / \  
{\rm O}(4) \times {\rm O}(20) \ .
\ee
We can think of the Grassmannian on the right
\be
{\rm O}(4,20) \  / \  {\rm O}(4) \times {\rm O}(20) 
\ee
as describing the choice of a positive-definite 4-dimensional subspace 
$\Pi \subset {\mathbb R}^{4,20}$, while the group on the left, ${\rm O}(\Gamma^{4,20})$,
leads to discrete identifications among them. Here ${\rm O}(\Gamma^{4,20})$ is
the group of isometries of a given fixed unimodular lattice 
$\Gamma^{4,20} \subset {\mathbb R}^{4,20}$.  (In physics terms, the lattice 
$\Gamma^{4,20}$ can be thought of as the D-brane charge lattice of string theory on K3.)
\medskip

Let  us denote by ${\cal H}^{(0)}\subset {\cal H}_{\rm RR}$ the subspace of 
(\ref{K3spec}) that consists of those RR states for which the right-moving states
are ground states. (Thus ${\cal H}^{(0)}$ consists of the states that contribute to
the elliptic genus.) ${\cal H}^{(0)}$ carries an action of the left-moving ${\cal N}=4$
superconformal algebra, and at any point in moduli space, its decomposition is of the 
form
\be\label{BPSdec}
{\cal H}^{(0)} = 20\cdot {\cal H}_{h=\frac{1}{4},j=0} \ \oplus  \ 
2 \cdot {\cal H}_{h=\frac{1}{4},j=\frac{1}{2}}
\ \oplus \ \bigoplus_{n=1}^{\infty} D_n \,  {\cal H}_{h=\frac{1}{4}+n,j=\frac{1}{2}} \ , 
\ee
where ${\cal H}_{h,j}$ denotes the irreducible ${\cal N}=4$ representation whose Virasoro
primary states have conformal dimension $h$ and transform in the spin $j$ representation 
of $\mathfrak{su}(2)$. The multiplicities $D_n$ are {\em not} constant over the moduli space,
but the above argument shows that 
\be
A_n = \hbox{Tr}_{D_n} (-1)^{\bar{F}} 
\ee
are (where $D_n$ is now understood not just as a multiplicity, but as a representation of 
the right-moving $(-1)^{\bar{F}}$ operator that determines the sign with which these states
contribute to the elliptic genus). In this language, the elliptic genus then takes the form
\be\label{ellgenN4}
\phi_{\rm K3}(\tau,z) = 20 \cdot \chi_{h=\frac{1}{4},j=0}(\tau,z) 
- 2 \cdot \chi_{h=\frac{1}{4},j=\frac{1}{2}}(\tau,z)
+ \sum_{n=1}^{\infty} A_n \cdot \chi_{h=\frac{1}{4}+n,j=\frac{1}{2}}(\tau,z) \ ,
\ee
where $\chi_{h,j}(\tau,z)$ is the `elliptic' genus of the corresponding ${\cal N}=4$
representation,
\be
\chi_{h,j} (\tau,z) = \hbox{Tr}_{{\cal H}_{h,j}} \bigl( q^{L_0-\frac{c}{24}}\, y^{J_0} (-1)^F \Bigr) \ ,
\ee
and we have used that $(-1)^{\bar{F}}$ takes the eigenvalues $+1$ and $-1$ on the 
$20$- and $2$-dimensional multiplicity spaces of the first two terms in (\ref{BPSdec}),
respectively. 

The key observation of Eguchi, Ooguri \& Tachikawa (EOT) \cite{EOT} was that the $A_n$ are
sums of dimensions of $\MM_{24}$ representation, in striking analogy to the original
Monstrous Moonshine conjecture of \cite{CN}; the first few terms are 
\begin{eqnarray}
A_1 & = & 90 = {\bf 45} + \overline{\bf 45} \\
A_2 & = & 462 = {\bf 231} + \overline{\bf 231} \\
A_3 & = & 1540 = {\bf 770} + \overline{\bf 770} \ ,
\end{eqnarray}
where ${\bf N}$ denotes a representation of $\MM_{24}$ of dimension $N$. 
Actually, they guessed correctly the first six coefficients; from $A_7$ onwards the guesses
become much more ambiguous (since the dimensions of the $\MM_{24}$ representations
are not that large) and they actually misidentified the seventh coefficient in their original
analysis. (We will come back to the question of why and how one can be certain about the 
`correct' decomposition shortly, see section~\ref{sec:decompose}.)
The alert reader will also notice that the first two coefficients in (\ref{BPSdec}), 
namely $20$ and $-2$, are not directly $\MM_{24}$ representations; the correct 
prescription is to introduce virtual representations and to write 
\be\label{remain}
20 = {\bf 23} - 3 \cdot {\bf 1} \ , \qquad - 2 = - 2 \cdot  {\bf 1} \ .
\ee
Recall that
$\MM_{24}$ is a sporadic finite simple group of order 
\be\label{Morder}
| \MM_{24} | = 2^{10} \cdot 3^3 \cdot 5 \cdot 7 \cdot 11 \cdot 23 = 244\, 823\, 040 \ .
\ee
It has $26$ conjugacy classes (which are denoted by 1A, 2A, 3A, $\ldots$, 23A, 23B, where
the number refers to the order of the corresponding group element) --- see eqs.\ 
(\ref{conj1}) and (\ref{conj2}) below for the full list --- and
therefore also $26$ irreducible representations whose dimensions range from 
$N=1$ to $N=10395$.  The Mathieu group $\MM_{24}$ can be defined as the subgroup of 
the permutation group $S_{24}$ that leaves the extended Golay code invariant; equivalently, 
it is the quotient of the automorphism group 
of the $\mathfrak{su}(2)^{24}$ Niemeier lattice, divided by the Weyl group. Thought of as a subgroup of 
$\MM_{24}\subset S_{24}$,  it contains the subgroup $\MM_{23}$ that is characterised 
by the condition that it leaves a given (fixed) element of $\{1,\ldots, 24\}$ invariant.

\subsection{Classical symmetries}

The appearance of  a Mathieu group in the elliptic genus of K3 is not totally surprising in view 
of the  Mukai theorem \cite{Mukai,Kondo}. It states that any finite group of symplectic automorphisms of a
K3 surface can be embedded into the Mathieu group 
$\MM_{23}$. The symplectic automorphisms of a K3 surface define symmetries
that act on the multiplicity spaces of the ${\cal N}=4$ representations, and therefore
explain part of the above findings. 
However, it is also clear from Mukai's argument that they do not even account for the 
full $\MM_{23}$ group. Indeed, every symplectomorphism of K3 has at least five orbits on 
the set $\{1,\ldots,24\}$, and thus not all elements of $\MM_{23}$ can be realised as a 
symplectomorphism.
More specifically,  of the $26$ conjugacy classes of $\MM_{24}$, 
$16$ have a representative in $\MM_{23}$, namely
\be\label{conj1}
\hbox{repr.\ in $\MM_{23}$:}\;\;\;\;\;
\begin{array}{ll}
\hbox{1A, 2A, 3A, 4B, 5A, 6A, 7A, 7B, 8A} \quad & \hbox{(geometric)} \\
\hbox{11A, 14A, 14B, 15A, 15B, 23A, 23B} \qquad &  \hbox{(non-geometric)\ ,}
\end{array}
\ee
where `geometric' means that a representative can be (and in fact is) realised by a 
geometric symplectomorphism (i.e.\ that the representative has at least five orbits when 
acting on the set 
$\{1,\ldots,24\}$), while `non-geometric' means that this is not the case. The remaining 
conjugacy classes do {\em not} have a representative in $\MM_{23}$, and are therefore 
not accounted for geometrically via the Mukai theorem
\be\label{conj2}
\hbox{no repr.\ in $\MM_{23}$:}\;\;\;\;\;
\hbox{2B, 3B, 4A, 4C, 6B, 10A, 12A, 12B, 21A, 21B} \ .
\ee
The classical symmetries can therefore only explain the symmetries in the first line
of (\ref{conj1}).  Thus an additional argument is needed in order to understand the origin 
of the other symmetries; we shall come back to this in section~\ref{sec:Theorem}.

\subsection{Evidence for Moonshine}\label{sec:decompose}

As was already alluded to above, in order to determine the `correct' decomposition
of the $A_n$ multiplicity spaces in terms of $\MM_{24}$ representations, we need
to study more than just the usual elliptic genus. By analogy with Monstrous Moonshine,
the natural objects to consider are the analogues of the McKay Thompson series 
\cite{Thompson1}. These are obtained from the elliptic genus upon replacing 
\be\label{guess}
A_n = \dim R_n \ \rightarrow \ \hbox{Tr}_{R_n} (g) \ ,
\ee
where $g\in \MM_{24}$, and $R_n$ is the $\MM_{24}$ representation whose dimension
equals the coefficient $A_n$; the resulting functions are then 
(compare (\ref{ellgenN4}))
\be\label{ellgentwinN4}
\phi_{g}(\tau,z) = \hbox{Tr}_{{\bf 23} - 3 \cdot {\bf 1}} (g) \, \chi_{h=\frac{1}{4},j=0}(\tau,z) 
- 2 \, \hbox{Tr}_{{\bf 1}}(g) \, \chi_{h=\frac{1}{4},j=\frac{1}{2}}(\tau,z)
+ \sum_{n=1}^{\infty} \hbox{Tr}_{R_n} (g)  \, \chi_{h=\frac{1}{4}+n,j=\frac{1}{2}}(\tau,z) \ .
\ee
The motivation for this definition comes from the observation
that if the underlying vector space ${\cal H}^{(0)}$, 
see eq.~(\ref{BPSdec}), of states contributing to the elliptic genus were 
to carry an action of $\MM_{24}$, $\phi_g(\tau,z)$ would equal the 
`twining elliptic genus', i.e.\ the elliptic genus twined by the action of $g$
\be\label{twining}
\phi_g(\tau,z) =  {\rm Tr}_{\rm {\cal H}^{(0)}} \Bigl(
g\, q^{L_0 - \frac{c}{24}} \, y^{J_0} \, (-1)^F \, 
\bar{q}^{\bar{L}_0 - \frac{\bar{c}}{24}}\, (-1)^{\bar{F}} \Bigr)  \ .
\ee
Obviously, a priori, it is not clear what the relevant $R_n$ in (\ref{guess}) are. However,
we have some partial information about them:
\begin{list}{(\roman{enumi})}{\usecounter{enumi}}
\item For any explicit realisation of a symmetry of a K3 sigma-model, we can 
calculate (\ref{twining}) directly. (In particular, for some symmetries, the relevant
twining genera had already been calculated in \cite{David:2006ji}.)
\item The observation of EOT determines the first six coefficients explicitly.
\item The twining genera must have special modular properties.
\end{list}
Let us elaborate on (iii). Assuming that the functions $\phi_g(\tau,z)$ have indeed an 
interpretation as in (\ref{twining}), they correspond in the usual orbifold notation of
string theory to the contribution 
\begin{eqnarray}\label{boxd}
\phi_g(\tau,z) \ \longleftrightarrow
& e\, &   \boxed{\phantom{{ \tiny \begin{array}{cc}
				\cdot & \cdot \hspace*{-0.2cm} \\  &  \hspace*{-0.2cm} \end{array}}}} \nonumber \\
& &  \hspace*{0.35cm}  g
\end{eqnarray}
where $e$ is the identity element of the group. Under a
modular transformation it is believed that these twining and twisted genera transform 
(up to a possible phase) as
\begin{equation}
\begin{array}{llllll}
& h & \boxed{\phantom{{\tiny \begin{array}{cc}
				\cdot & \cdot \hspace*{-0.2cm} \\  &  \hspace*{-0.2cm} \end{array}}}}
\hspace*{0.5cm}
& \stackrel{\left(\begin{array}{cc} a & b \\ c & d \end{array}\right)} {\xrightarrow{\hspace*{2cm}}} \hspace*{0.5cm}
& h^d g^c & \boxed{\phantom{{\tiny \begin{array}{cc}
				\cdot & \cdot \hspace*{-0.2cm} \\  &  \hspace*{-0.2cm}\end{array}}}}  \\
& & \hspace*{0.3cm}  g
& & &   \hspace*{0.05cm} g^a h^b
\end{array}
\end{equation}
The twining genera (\ref{boxd}) are therefore invariant (possibly up to a phase)
under the modular transformations with
\begin{equation}\label{cons}
\gcd(a,o(g))=1  \qquad \hbox{and} \qquad c = 0 \ \ \hbox{mod $o(g)$,} 
\end{equation}
where $o(g)$ is the order of the group element $g$ and we used that for 
$\gcd(a,o(g))=1$, the group element $g^a$ is in the same conjugacy class as 
$g$ or $g^{-1}$. (Because of reality, the twining genus of $g$ and $g^{-1}$ should
be the same.) Since $ad-bc=1$, the second 
condition implies the first, and we thus conclude that $\phi_g(\tau,z)$ should be 
(up to a possible multiplier system) a weak Jacobi form of weight zero and index one 
under the subgroup of  ${\rm SL}(2,{\mathbb Z})$ 
\begin{equation}
\Gamma_0(N) = \left\{
\left( \begin{array}{cc} a & b \\ c & d \end{array} \right) \in
{\rm SL}(2,{\mathbb Z}) \ : \  c = 0 \,\, \hbox{mod $(N)$} \right\}  \ ,
\end{equation}
where $N=o(g)$. This is a relatively strong condition, and knowing the first few terms 
(for a fixed multiplier system)  determines the function uniquely. In order to use
this constraint, however, it is important to know the multiplier system. An ansatz (that
seems to work, see below) was made in \cite{Gaberdiel:2010ca}
\be\label{conj} \phi_g\Bigl(\frac{a\tau+b}{c\tau+d},\frac{z}{c\tau+d}\Bigr)
=e^{\frac{2\pi i cd}{Nh}}\,e^{\frac{2\pi i\, cz^2}{c\tau+d}}\,\phi_g(\tau,z)\ ,\qquad
\begin{pmatrix}a & b\\ c&d \end{pmatrix}\in \Gamma_0(N)\ ,
\ee
where $N$ is again the order of $g$ and $h|\gcd(N,12)$. The multiplier system is
trivial ($h=1$) if and only if $g$ contains a representative in $\MM_{23}\subset \MM_{24}$.
For the other conjugacy classes, the values are tabulated in table~\ref{t:hvalues}.
\begin{table}[htb]\centerline{
$\begin{array}{|c|ccccccccc|}\hline
\text{Class}  & {\rm 2B} & {\rm 3B} & {\rm 4A} & {\rm 4C} & {\rm 6B}
& {\rm 10A} & {\rm 12A} & {\rm 12B} & {\rm 21AB} \\ \hline
h  & 2 & 3 & 2 & 4 & 6 & 2 & 2 & 12 & 3\\ \hline
\end{array}$}\caption{Value of $h$ for the conjugacy classes in (2.28).}\label{t:hvalues}
\end{table}
It was noted in \cite{CD1} that $h$ equals the length of the shortest cycle
(when interpreted as a permutation in $S_{24}$, see table~1 of   \cite{CD1}).

Using this ansatz, explicit expressions for all twining genera were determined in 
\cite{Gaberdiel:2010ca}; independently, the same twining genera were also found 
(using guesses based on the cycle shapes of the corresponding $S_{24}$ representations) 
in  \cite{Eguchi:2010fg}. (Earlier partial results had been obtained in 
\cite{Cheng:2010pq} and \cite{Gaberdiel:2010ch}.) 

These explicit expressions for the twining genera then allow for a very non-trivial
check of the EOT proposal. As is clear from their definition in (\ref{ellgentwinN4}), 
they determine the coefficients
\be
\hbox{Tr}_{R_n} (g) \qquad \hbox{for all $g\in\MM_{24}$ and all $n\geq 1$.}
\ee
This information is therefore sufficient to {\em determine} the representations $R_n$,
i.e.\ to calculate their decomposition into irreducible $\MM_{24}$ 
representations, for all $n$. We have
worked out the decompositions explicitly for the first $500$ coefficients, and we have found that
each $R_n$ can be written as a direct sum of $\MM_{24}$ representations with
non-negative integer multiplicities \cite{Gaberdiel:2010ca}. (Subsequently  
\cite{Eguchi:2010fg} tested this property for the first $600$ coefficients, and apparently
Tachikawa has also checked it for the first $1000$ coefficients.) 
Terry Gannon has informed us that this information is sufficient to prove that
the same will then happen for all $n$ \cite{Gannon}. In some sense this then proves
the EOT conjecture.

\section{Symmetries of K3 models}\label{sec:Theorem}

While the above considerations establish in some sense the EOT conjecture, they
do not offer any insight into why the elliptic genus of K3 should exhibit
an $\MM_{24}$ symmetry. This is somewhat similar to the original situation in Monstrous
Moonshine, after Conway and Norton had found the various Hauptmodules by
somewhat similar techniques. Obviously, in the case of Monstrous Moonshine,
many of these observations were subsequently explained by the construction of the
Monster CFT (that possesses the Monster group as its
automorphism group) \cite{Borcherds,FLM}. So we should similarly ask for
a microscopic explanation of these findings. 

In some sense it is clear what the analogue of the Monster CFT in the current context
should be: we know that the function in question {\em is} the elliptic genus of
K3. However, there is one problem with this. As we mentioned before, there is not
just one K3 sigma-model, but rather a whole moduli space (see eq.\ (\ref{moduli}))
of such CFTs. So the simplest explanation of the EOT observation would be if there 
is (at least) one special K3 sigma-model that has $\MM_{24}$ as its automorphism group.
Actually, the relevant symmetry group should commute with the action of the 
${\cal N}=(4,4)$ superconformal symmetry (since it should act on the multiplicity
spaces in ${\cal H}^{(0)}$, see eq.\ (\ref{BPSdec})). Furthermore, since the two
${\cal N}=4$ representations with $h=\tfrac{1}{4}$ and $j=\tfrac{1}{2}$ are singlets --- recall
that the coefficient $-2$ transforms as 
$- 2=- 2\cdot {\bf 1}$, see (\ref{remain}) --- the automorphism must act trivially
on the 4 RR ground states that transform in the $({\bf 2},{\bf 2})$ representation of the 
$\mathfrak{su}(2)_L \times \mathfrak{su}(2)_R$ subalgebra of ${\cal N}=(4,4)$. Note that
these four states generate the simultaneous half-unit spectral flows in the left- 
and the right-moving sector; the requirement that the symmetry leaves them invariant
therefore means that spacetime supersymmetry is preserved.

Recall from (\ref{moduli}) that the different K3 sigma-models are parametrised
by the choice of a positive-definite $4$-dimensional
subspace $\Pi\subset {\mathbb R}^{4,20}$, modulo some
discrete identifications. Let us denote by $G_\Pi$ the group of symmetries of 
the sigma-model described by $\Pi$ that commute with the action of 
${\cal N}=(4,4)$ and preserve the RR ground states in the $({\bf 2},{\bf 2})$ (see above). 
It was argued in  \cite{Gaberdiel:2011fg} that $G_\Pi$ is precisely the 
subgroup of ${\rm O}(\Gamma^{4,20})$ consisting of those elements that leave 
$\Pi$ pointwised fixed. The possible symmetry groups $G_\Pi$ can then be classified
following essentially the paradigm of the Mukai-Kondo argument for the
symplectomorphisms of K3 surfaces \cite{Mukai,Kondo}. The outcome of 
the analysis can be summarised by the following theorem  \cite{Gaberdiel:2011fg}:
\smallskip

\noindent {\bf Theorem:}
{\it Let $G$ be the group of symmetries of a non-linear sigma-model on $K3$ preserving 
the $\N=(4,4)$ superconformal algebra as well as the spectral flow operators. 
One of the following possibilities holds:
\begin{list}{{\rm (\roman{enumi})}}{\usecounter{enumi}}
\item $G=G'. G''$, where $G'$ is a subgroup of $\ZZ_2^{11}$, and $G''$ is a subgroup of  
$\MM_{24}$ with at least four orbits when acting as a permutation on
$\{1,\ldots,24\}$
\item $G=5^{1+2}:\ZZ_4$
\item $G=\ZZ_3^4:A_6$
\item $G=3^{1+4}:\ZZ_2.G''$, where $G''$ is either trivial, $\ZZ_2$ or $\ZZ_2^2$.
\end{list}}

\noindent 
Here $G=N.Q$ means that $N$ is a normal subgroup of $G$, and $G/N\cong Q$; when $G$ is the semidirect product of $N$ and $Q$, we denote it by $N:Q$.
Furthermore, $p^{1+2n}$ is an extra-special group of order $p^{1+2n}$, which
is an extension of $\mathbb{Z}_p^{2n}$ by a central element of order $p$. 
\smallskip

We will give a sketch of the proof below (see section~\ref{sec:proof}), but for the
moment let us comment on the implications of this result. First of all, our initial
expectation from above is not realised: none of these groups $G\equiv G_\Pi$ contains 
$\MM_{24}$. In particular, the twining genera for the conjugacy classes 
12B, 21A, 21B, 23A, 23B of $\MM_{24}$ cannot be realised by any symmetry of a 
K3 sigma-model. Thus we cannot give a direct explanation of the EOT observation
along these lines.

Given that the elliptic genus is constant over the moduli space, one may then hope
that we can explain the origin of $\MM_{24}$ by `combining' symmetries from different
points in the moduli space. As we have mentioned before, this is also similar to what 
happens for the geometric symplectomorphisms of K3:  it follows from the Mukai
theorem that the Mathieu group $\MM_{23}$ is the smallest group 
that contains all symplectomorphisms, but there is no K3 surface where all of 
$\MM_{23}$ is realised, and indeed, certain generators of $\MM_{23}$ can never be 
symmetries, see (\ref{conj1}). However, also this explanation of the EOT 
observation is somewhat problematic: as is clear from the above theorem, not all symmetry groups 
of  K3 sigma-models are in fact subgroups of $\MM_{24}$. In particular, none
of the cases (ii), (iii) and (iv) (as well as case (i) with $G'$ non-trivial) 
have this property, as can be easily seen by comparing
the prime factor decompositions of their orders to (\ref{Morder}). The smallest
group that contains all groups of the theorem is the Conway group ${\rm Co}_1$, but
as far as we are aware, there is no evidence of any `Conway Moonshine' in 
the elliptic genus of K3. 

One might speculate  that, generically, the group $G$ must be a subgroup of $\MM_{24}$, 
and that the models whose symmetry group is not contained in $\MM_{24}$
are, in some sense, special or `exceptional' points in the moduli space.
%; from now on we shall call these symmetry groups as well as the corresponding models 
%`exceptional'.
 In order to make this idea precise, it is useful
to analyse the exceptional models in detail. In \cite{Gaberdiel:2011fg}, some examples 
have been provided of case (i) with non-trivial $G'$ (a torus orbifold $\mathbb{T}^4/\ZZ_2$ 
or the Gepner model $2^4$, believed to be equivalent to a $\TT^4/\ZZ_4$ orbifold), 
and of case (iii) (the Gepner 
model $1^6$, which is believed to be equivalent to a $\TT^4/\mathbb{Z}_3$ orbifold,
see also \cite{Fluder}). For the cases (ii) and (iv), only an existence proof was given. 
In section~\ref{sec:Z5}, we will improve the situation by constructing 
in detail an example of case (ii), 
realised as  an asymmetric $\ZZ_5$-orbifold of a torus $\mathbb{T}^4$. Furthermore, in 
section~\ref{sec:Z3} we will briefly discuss the $\ZZ_3$-orbifold of a torus and the 
explicit realisation of its symmetry group, corresponding to cases (ii) and (iv) for any $G''$.  

Notice that all the examples of exceptional models known so far are provided by torus 
orbifolds. In fact, we will show below (see section~\ref{sec:symme}) that all cyclic torus orbifolds have 
exceptional symmetry groups. Conversely, we will prove that the cases (ii)--(iv) of the theorem 
are always realised by (cyclic) torus orbifolds. On the other hand, as we shall also explain,
some of the exceptional models in case (i) are not cyclic torus orbifolds.

%These considerations suggest that the special cases may 
%all be realised by torus orbifolds. 
%K3 $\sigma$-models that can be constructed as torus orbifolds
%$\mathbb{T}^4/\langle g \rangle$ have the special property
%that they possess a symmetry $\tilde{g}$, the so-called `quantum symmetry' 
%associated to $g$, such that 
%\be
%\hbox{K3} / \langle \tilde{g} \rangle \cong \mathbb{T}^4 \ .
%\ee
%They may therefore play a similar role to the special group elements in
%Monstrous Moonshine that lead, after orbifolding, to the Leech theory (rather
%than the Monster CFT), see \cite{Tuite:1992tt,Tuite:1993hy}. As was noted there, 
%the corresponding  McKay-Thompson series are all non-Fricke, and hence 
%behave less well than in the other cases. It would be very interesting to understand 
%in more  detail what the precise situation in the context of Mathieu Moonshine is; we 
%hope to come back to these questions elsewhere \cite{GV}.

\subsection{Sketch of the proof of the Theorem}\label{sec:proof}

In this subsection, we will describe the main steps in the proof of the above theorem;
the details can be found in \cite{Gaberdiel:2011fg}.

It was argued in \cite{Gaberdiel:2011fg} that the 
supersymmetry preserving automorphisms of the non-linear sigma-model
characterised by $\Pi$ generate the group $G\equiv G_\Pi$ that consists of 
those elements of $O(\Gamma^{4,20})$ that leave $\Pi$ pointwise fixed.  
Let us denote by $L^G$ the sublattice of $G$-invariant vectors of  
$L\equiv \Gamma^{4,20}$, and define $L_G$ to be its orthogonal complement
that carries a genuine action of $G$. Since $L^G\otimes \RR$ contains the subspace 
$\Pi$, it follows that $L^G$ has signature $(4,d)$ for some $d\ge 0,$ so that $L_G$ is a 
negative-definite lattice of rank $20-d$. In \cite{Gaberdiel:2011fg}, it is proved that, 
for any consistent model, $L_G$ can be embedded (up to a change of sign in its quadratic form) 
into the Leech lattice $\Lambda$, the unique $24$-dimensional positive-definite even 
unimodular lattice containing no vectors of squared norm $2$. Furthermore, the action of 
$G$ on $L_G$ can be extended to an action on the whole of $\Lambda$, such that the 
sublattice $\Lambda^G\subset\Lambda $ of vectors fixed by $G$ is the orthogonal 
complement of $L_G$ in $\Lambda$. This construction implies that $G$ must be a subgroup 
of ${\rm Co}_0\equiv {\rm Aut}(\Lambda)$ that fixes a sublattice $\Lambda^G$ of rank 
$4+d$. Conversely, it can be shown that any such subgroup of ${\rm Aut}(\Lambda)$ is the 
symmetry group of some K3 sigma-model.
\medskip

This leaves us with characterising the possible subgroups of the finite group
${\rm Co}_0\equiv {\rm Aut}(\Lambda)$ that stabilise a suitable sublattice; problems of
this kind have been studied in the mathematical literature before. 
In particular, the stabilisers of sublattices of rank at least $4$ are, generically, the 
subgroups of $\ZZ_2^{11}:\MM_{24}$ described in case (i) of the theorem above.
The three cases (ii), (iii), (iv) arise when the invariant sublattice $\Lambda^G$ is 
contained in some $\mathcal{S}$-lattice $S\subset\Lambda$. An $\mathcal{S}$-lattice $S$ 
is a primitive sublattice of $\Lambda$ such that each vector of $S$ is congruent modulo $2S$ to a 
vector of norm $0$, $4$ or $6$. Up to isomorphisms, there are only three kind of $\mathcal{S}$-lattices 
of rank at least four; their properties are summarised in the following table:
$$\begin{array}{ccccc}
    \text{Name} \qquad&\text{type}\qquad & \rk S \qquad & {\rm Stab}(S) \qquad & {\rm Aut}(S)\\
    (A_2\oplus A_2)'(3)  \qquad& 2^93^6 \qquad & 4 \qquad & \ZZ_3^4: A_6\qquad  & \ZZ_2\times(S_3\times S_3).\ZZ_2\\
    A_4^*(5)  \qquad& 2^{5}3^{10} \qquad & 4 \qquad & 5^{1+2}:\ZZ_4 \qquad & \ZZ_2\times S_5\\
    E_6^*(3)  \qquad& 2^{27}3^{36} \qquad  & 6 \qquad & 3^{1+4}:\ZZ_2 \qquad & \ZZ_2\times W(E_6) \ .
\end{array}
$$
Here, $S$ is characterised by the type $2^p3^q$, which indicates that $S$ contains 
$p$ pairs of opposite vectors of norm $4$ (type 2) and $q$ pairs of opposite vectors of norm 
$6$ (type 3). The group ${\rm Stab}(S)$ is the pointwise stabiliser of $S$ in ${\rm Co}_0$ 
and ${\rm Aut}(S)$ is the quotient of the setwise stabiliser of $S$ modulo its pointwise 
stabiliser ${\rm Stab}(S)$. The group ${\rm Aut}(S)$ always contains a central $\ZZ_2$ 
subgroup, generated by the transformation that inverts the sign of all vectors of the Leech lattice. 
The lattice of type $2^{27}3^{36}$ is isomorphic to the weight 
lattice (the dual of the root lattice) of $E_6$ with quadratic form rescaled by $3$ (i.e.\ 
the roots in $E_6^*(3)$ have squared norm $6$), and ${\rm Aut}(S)/\ZZ_2$ is isomorphic to the 
Weyl group $W(E_6)$ of $E_6$. Similarly, the lattice of type $2^{5}3^{10}$ is the weight 
lattice of $A_4$ rescaled by $5$, and ${\rm Aut}(S)/\ZZ_2$ is isomorphic to the Weyl group
$W(A_4)\cong S_5$ of $A_4$. Finally, the type $2^93^6$ is the $3$-rescaling of a lattice 
$(A_2\oplus A_2)'$ obtained by adjoining to 
the root lattice $A_2\oplus A_2$ an element $(e_1^*,e_2^*)\in A_2^*\oplus A_2^*$, with $e_1^*$ 
and $e_2^*$ of norm $2/3$. The latter ${\mathcal S}$-lattice can also be described as the 
sublattice of vectors of $E_6^*(3)$ that are orthogonal to an $A_2(3)$ sublattice of $E_6^*(3)$.  
The group ${\rm Aut}(S)/\ZZ_2$ is 
the product $(S_3\times S_3).\ZZ_2$ of the Weyl groups $W(A_2)=S_3$, and the
$\ZZ_2$ symmetry that exchanges the two $A_2$ and maps $e_1^*$ to $e_2^*$.

\smallskip

The cases (ii)--(iv) of the above theorem correspond to $\Lambda^G$ being 
isomorphic to $A_4^*(5)$ (case ii), to $(A_2\oplus A_2)'(3)$ (case iii) or to a sublattice of 
$E_6^*(3)$ different from $(A_2\oplus A_2)'(3)$ (case iv). In the cases (ii) and (iii), 
$G$ is isomorphic to ${\rm Stab}(S)$. In case (iv), ${\rm Stab}(S)$ is, generically, a normal 
subgroup of $G$, and $G''\cong G/{\rm Stab}(S)$ is a subgroup of 
${\rm Aut}(S)\cong \ZZ_2\times W(E_6)$ that fixes a sublattice $\Lambda^G\subseteq E_6^*(3)$, 
with $\Lambda^G\neq (A_2\oplus A_2)'(3)$, of rank at least $4$. The only non-trivial 
subgroups of $\ZZ_2\times W(E_6)$ with these properties are $G''=\ZZ_2$, which corresponds to 
$\Lambda^G$ being orthogonal to a single vector of norm $6$ in $E_6^*(3)$ (a rescaled root), 
and $G''=\ZZ_2^2$, which corresponds to $\Lambda^G$ being orthogonal to two orthogonal 
vectors of norm $6$.\footnote{The possibility $G''=\ZZ_4$ that has been considered in 
\cite{Gaberdiel:2011fg} has to be excluded, since there are no elements of order $4$ in 
$W(E_6)$ that preserve a four-dimensional sublattice of $E_6^*(3)$.} If $\Lambda^G$ is 
orthogonal to two vectors $v_1,v_2\in E_6^*(3)$ of norm $6$, with $v_1\cdot v_2=-3$, 
then $\Lambda^G\cong (A_2\oplus A_2)'(3)$ and case (iii) applies. 

\section{Symmetry groups of torus orbifolds}\label{sec:symme}

In this section we will prove that all K3 sigma-models that are realised as 
(possibly left-right asymmetric) orbifolds of $\TT^4$ by a cyclic group have an 
`exceptional' group of symmetries, i.e.\ their symmetries are not a subgroup of $\mathbb{M}_{24}$.
Furthermore, these torus orbifolds account for most of the exceptional models (in 
particular, for all models in the cases (ii)--(iv) of the theorem). On the other hand,
as we shall also explain, there are exceptional models in case (i) that are not 
cyclic torus orbifolds.

Our reasoning is somewhat reminiscent of the construction of \cite{Tuite:1992tt,Tuite:1993hy} in the 
context of Monstrous Moonshine. Any $\ZZ_n$-orbifold of a conformal field theory has
an automorphism $g$ of order $n$, called the \emph{quantum symmetry}, which acts trivially on the 
untwisted sector and by multiplication by the phase $\exp(\frac{2\pi i k}{n})$ on the 
$k$-th twisted sector. Furthermore, the orbifold of the orbifold theory by the group generated 
by the  quantum symmetry $g$, is equivalent to the original conformal field theory 
\cite{Ginsparg:1988ui}.  This observation is the key for characterising K3 models that 
can be realised as torus orbifolds:

\bigskip

{\sl A K3 model $\C$ is a $\ZZ_n$-orbifold of a torus model if and only if it has a symmetry $g$  of 
order $n$ such that $\C/\langle g\rangle$ is a consistent orbifold
equivalent to a torus model.}

\bigskip

%Let us consider a K3 $\sigma$-model $\C$ with some symmetry $g$ of order $n$, and let 
%$\tilde \C$ be the orbifold $\C/\langle g\rangle$. Consistency of the orbifold construction 
%requires that level-matching is satisfied. This is equivalent to the condition that the 
%twining genus $\phi_g(\tau,z)$ is a Jacobi form for $\Gamma_0(o(g))$ with trivial multiplier system. 
%Let us assume that this condition is satisfied, so that the orbifold theory $\tilde \C$ is a consistent 
%$\N=(4,4)$ SCFT with central charge $c=6$.\footnote{For the case when $g$ is the quantum
%symmetry of a K3 orbifold, this condition will be automatically satisfied.}
%Any such theory is believed to admit a description as a non-linear $\sigma$-model with 
%target space either a torus $\TT^4$ or a K3 manifold, and the condition that the 
%`target space' is $\TT^4$ is simply that the elliptic genus of the orbifold theory 
%$\tilde{C}$ vanishes.
%% \footnote{Actually, our results can be rephrased in a way that is independent of this assumption. In fact, it can be proved that there are two classes of $\N=(4,4)$ SCFT at $c=6$, distinguished by their elliptic genus; one of them contains all K3 models and the other all torus models.}.  
%Actually, since the space of weak Jacobi forms of weight zero and index one is one-dimensional,
%this is equivalent to the requirement that the elliptic genus $\tilde\phi(\tau,z)$ of $\tilde \C$ 
%vanishes at $z=0$. (Geometrically, this simply means that the Euler number of the `target space'
%vanishes.) 

In order to see this, suppose that $\C_{K3}$ is a K3 sigma-model that can be 
realised as a torus orbifold
$\C_{K3} = \tilde\C_{\TT^4} /\langle \tilde{g} \rangle$, where $\tilde{g}$ is a symmetry of order $n$ of 
the torus model $\tilde \C_{\TT^4}$. Then $\C_{K3}$ 
possesses a `quantum symmetry' $g$ of order $n$, such that the orbifold of $\C_{K3}$ by 
$g$ describes again the original torus model, $\tilde{\C}_{\TT^4} = \C_{K3} / \langle g \rangle$. 

Conversely, suppose $\C_{K3}$ has a symmetry $g$ of order $n$, such that the orbifold
of $\C_{K3}$ by $g$ is consistent, i.e.\ satisfies the level matching condition --- this is the
case if and only if the twining genus $\phi_g$ has a trivial multiplier system --- 
and leads to a torus model $\C_{K3} / \langle g \rangle = \tilde{\C}_{\mathbb{T}^4}$. Then 
$\C_{K3}$ itself is a torus orbifold
since we can take the orbifold of $\tilde\C_{\mathbb{T}^4}$ by the quantum symmetry associated
to $g$, and this will, by construction, lead back to $\C_{K3}$. 
\smallskip

Thus we conclude that $\C\equiv \C_{K3}$ can be realised as a torus orbifold if and only if $\C$ contains
a symmetry $g$ such that (i) $\phi_{g}$ has a trivial multiplier system; 
and (ii) the orbifold of $\C$ by $g$ leads to a torus model $\tilde\C_{\TT^4}$. It is believed that
the orbifold of $\C$ by any $\N=(4,4)$-preserving symmetry group, if consistent, will describe a sigma-model with target
space either a torus $\TT^4$ or a K3 manifold. The two cases can be distinguished
by calculating the elliptic genus; in particular, if the target space is a torus, 
the elliptic genus vanishes. Actually, since the space of weak Jacobi forms of weight zero 
and index one is 1-dimensional, this condition is equivalent to the requirement that the 
elliptic genus $\tilde\phi(\tau,z)$ of $\tilde \C = \C / \langle g \rangle$ vanishes at $z=0$.

Next we recall that the elliptic genus of the orbifold by a group element $g$ of order $n=o(g)$  
is given by the usual orbifold formula
\be 
\tilde\phi(\tau,z)=\frac{1}{n}\sum_{i,j=1}^{n} \phi_{g^i,g^j}(\tau,z)\ ,
\ee 
where $\phi_{g^i,g^j}(\tau,z)$ is the twining genus for $g^j$ in the $g^i$-twisted sector; this
can be obtained by a modular transformation from the untwisted twining genus 
$\phi_{g^d}(\tau,z)$ with $d=\gcd(i,j,n)$. As we have explained above,
it is enough to evaluate the elliptic genus for $z=0$. Then
\be\label{const}
\phi_{g^d}(\tau,z=0)=\Tr_{\bf 24}(g^d)\ ,
\ee 
where $\Tr_{\bf 24}(g^d)$ is the trace of $g^d$ over the $24$-dimensional space of 
RR ground states, and since (\ref{const}) is constant (and hence modular invariant) 
we conclude that 
\be\label{elliptorbif} 
\tilde\phi(\tau,0)=\frac{1}{n}\sum_{i,j=1}^{n} \Tr_{\bf 24}(g^{\gcd(i,j,n)})\ .
\ee
According to the theorem in section~\ref{sec:Theorem}, all symmetry groups of K3 
sigma-models are subgroups 
of  ${\rm Co}_0$ and, in fact, $\Tr_{\bf 24}(g^d)$ coincides with the trace of 
$g^d\in {\rm Co}_0$ in the standard $24$-dimensional representation of ${\rm Co}_0$. 
Thus, the elliptic genus of the orbifold model $\tilde \C=\C/\langle g\rangle$ only depends on the 
conjugacy class of $g$ in ${\rm Co}_0$. The group ${\rm Co}_0$ contains $167$ conjugacy 
classes, but only $42$ of them contain symmetries that are realised by some 
K3 sigma-model, i.e.\ 
elements that fix at least a four-dimensional subspace in the standard $24$-dimensional 
representation of ${\rm Co}_0$. If $\Tr_{\bf 24}(g)\neq 0$ (this happens for $31$ of the above 
$42$ conjugacy classes), the twining genus $\phi_g(\tau,z)$ has necessarily a trivial multiplier 
system, and the orbifold $\C/\langle g\rangle$ is consistent. These classes are listed in 
the following table, together with the dimension of the space that is fixed by $g$, the trace over 
the  $24$-dimensional representation, and the elliptic genus $\tilde\phi(\tau,z=0)$ of the 
orbifold model $\tilde \C$ (we underline the classes that restrict to $\MM_{24}$ conjugacy 
classes):

{\small
$$
\begin{array}{c|ccccccccccccccccc}
\text{${\rm Co}_0$-class} &   \text{\underline{1A}} & \text{\underline{2B}} & \text{2C} & \text{\underline{3B}} & \text{3C} & \text{4B} & \text{\underline{4E}} &
\text{4F} & \text{\underline{5B}} & \text{5C} & \text{6G} & \text{6H} & \text{6I} &
\text{\underline{6K}} &
\text{6L} & \text{6M} & \text{\underline{7B}} \\
 \text{dim fix} & 24 & 16 & 8 & 12 & 6 & 8 & 10 & 6 & 8 & 4 & 6 & 6 & 6 & 8 & 4 & 4 & 6 \\
 \Tr_{\bf 24}(g) &  24 & 8 & -8 & 6 & -3 & 8 & 4 & -4 & 4 & -1 & -4 & 4 & 5 & 2 & -2 & -1 & 3 \\
 \tilde\phi(\tau,0) &24 & 24 & 0 & 24 & 0 & 24 & 24 & 0 & 24 & 0 & 0 & 24 & 24 & 24 & 0 & 0 & 24
\end{array}
$$
$$
\begin{array}{c|cccccccccccccc}
 \text{${\rm Co}_0$-class} & \text{8D} & \text{\underline{8G}} & \text{8H} & \text{9C} & \text{10F} & \text{10G} & \text{10H} &
   \text{\underline{11A}} & \text{12I} & \text{12L} & \text{12N} & \text{12O} & \text{\underline{14C}} &
   \text{\underline{15D}} \\
 \text{dim fix} & 4 & 6 & 4 & 4 & 4 & 4 & 4 & 4 & 4 & 4 & 4 & 4 & 4 & 4 \\
 \Tr_{\bf 24}(g) &  4 & 2 & -2 & 3 & -2 & 2 & 3 & 2 & 2 & 1 & -2 & 2 & 1 & 1 \\
 \tilde\phi(\tau,0) & 24 & 24 & 0 & 24 & 0 & 24 & 24 & 24 & 24 & 24 & 0 & 24 & 24 & 24
\end{array}
$$}

\bigskip

Note that the elliptic genus of the orbifold theory $\tilde\C$ is always $0$ or $24$, 
corresponding to a torus or a K3 sigma-model, respectively.
Out of curiosity, we have also computed the putative elliptic genus $\tilde\phi(\tau,0)$ 
for the $11$ classes of symmetries $g$ with $\Tr_{24}(g)=0$ for which we do not 
expect the orbifold to make sense --- the corresponding twining genus $\phi_g$ will
typically have a non-trivial multiplier system, and hence the orbifold will not satisfy 
level-matching.
Indeed, for almost none of these cases is $\tilde\phi(\tau,0)$ equal to $0$ or $24$, thus 
signaling an inconsistency of the orbifold model:
$$\begin{array}{c|ccccccccccc}
 \text{${\rm Co}_0$-class} & \text{\underline{2D}} & \text{\underline{3D}} & \text{4D} & \text{\underline{4G}} & \text{\underline{4H}} & \text{6O} & \text{\underline{6P}} &
   \text{8C} & \text{8I} & \text{\underline{10J}} & \text{\underline{12P}} \\
 \text{dim fix} & 12 & 8 & 4 & 8 & 6 & 6 & 4 & 4 & 4 & 4 & 4 \\
 \Tr_{\bf 24}(g) & 0 & 0 & 0 & 0 & 0 & 0 & 0 & 0 & 0 & 0 & 0 \\
 \tilde\phi(\tau,0) &12 & 8 & 0 & 12 & 6 & 12 & 4 & 12 & 6 & 12 & 12
\end{array}
$$
The only exception is the class ${\rm 4D}$, which might define a consistent orbifold 
(a torus model). It follows that a K3 model $\C$ is the $\ZZ_n$-orbifold of a torus model 
if and only if it contains a symmetry $g$ in one of the classes
\begin{align}\label{qusymm} 
&{\rm 2C},\ {\rm 3C},\ {\rm 4F},\ {\rm 5C},\ {\rm 6G},\ {\rm 6L},\ {\rm 6M},\ {\rm 8H},\ 
{\rm 10F},\ {\rm 12N},\\
&{\rm 4B},\ {\rm 4D}, \ {\rm 6H},\ {\rm 6I},\ {\rm 8C},\ {\rm 8D},\  {\rm 9C},\ 
{\rm 10G},\ {\rm 10H},\  {\rm12I},\  {\rm 12L},\  {\rm 12O} \notag
\end{align} 
of ${\rm Co}_0$.\footnote{We should emphasise that for us  the term `orbifold' always refers 
to a conformal field theory (rather than a geometrical) construction. 
Although a non-linear sigma-model on a geometric orbifold $M/\ZZ_N$ always 
admits an interpretation as a CFT orbifold, the converse is not always true. In particular,
there exist asymmetric orbifold constructions that do not have a direct geometric 
interpretation, see for example section~\ref{sec:Z5}.} 
Here we have also included (in the second line) those classes of elements
$g\in {\rm Co}_0$ for which $\C/\langle g^i\rangle$ is a torus model, for some power $i>1$.
Our main observation is now that none of the ${\rm Co}_0$ classes in (\ref{qusymm})
restricts to a class in $\MM_{24}$, i.e.\
\bigskip

{\sl All K3 models that are realised as $\ZZ_n$-orbifolds of torus models are exceptional. 
In particular, the quantum symmetry is not an element of $\MM_{24}$.}

\bigskip

One might ask whether the converse is also true, i.e.\ whether all exceptional models 
are cyclic  torus orbifolds. This is not quite 
the case: for example, the classification theorem of section~\ref{sec:Theorem} 
predicts the existence of models with a symmetry group $G\cong GL_2(3)$ (the group of 
$2\times 2$ invertible matrices on the field $\FF_3$ with $3$ elements). The group $G$ 
contains no elements in the classes \eqref{qusymm}, so the model is not a cyclic torus orbifold; 
on the other hand, $G$ contains elements in the class ${\rm 8I}$ of ${\rm Co}_0$, which 
does not restrict to $\MM_{24}$. A second counterexample is a family of models with a 
symmetry $g$ in the class ${\rm 6O}$ of ${\rm Co}_0$. A generic point of this family is not 
a cyclic torus model (although some special points are), since the full symmetry group is 
generated by $g$ and contains no elements in \eqref{qusymm}. Both these counterexamples 
belong to case (i) of the general classification theorem. In fact, we can prove that 

\bigskip

{\sl The symmetry group $G$ of a K3 sigma-model $\C$ contains a subgroup 
$3^{1+4}:\ZZ_2$ (cases (iii) and (iv) of the theorem) if and only if $\C$ is a 
$\ZZ_3$-orbifold of a torus model. Furthermore, $G=5^{1+2}:\ZZ_4$ (case (ii)) if and 
only if $\C$ is a $\ZZ_5$-orbifold of a torus model.
}

\bigskip

The proof goes as follows. All subgroups of ${\rm Co}_0$ of the form 
$3^{1+4}:\ZZ_2$ (respectively, $5^{1+2}:\ZZ_4$) contain an element in the class 
3C (resp., 5C), and therefore the corresponding models are $\ZZ_3$ (resp., $\ZZ_5$) 
torus orbifolds. Conversely, consider a $\ZZ_3$-orbifold of a torus model. Its symmetry group
$G$ contains the quantum symmetry $g$ in class 3C of ${\rm Co}_0$. (It must contain
a symmetry generator of order three whose orbifold leads to a torus, and 3C is then
the only possibility.) The sublattice 
$\Lambda^{\langle g\rangle}\subset\Lambda$ fixed by $g$ is the 
$\mathcal{S}$-lattice $2^{27}3^{36}$ \cite{Curtis}. From the classification theorem, 
we know that $G$ is the stabiliser of a sublattice $\Lambda^G\subset\Lambda$ of rank at 
least $4$. Since $\Lambda^G\subseteq\Lambda^{\langle g\rangle}$, $G$ contains 
as a subgroup the stabiliser of $\Lambda^{\langle g\rangle}$, namely $3^{1+4}:\ZZ_2$. 

Analogously, a $\ZZ_5$ torus orbifold always has a symmetry in class 5C, whose fixed 
sublattice $\Lambda^{\langle g\rangle}$ is the $\mathcal{S}$-lattice $2^53^{10}$ \cite{Curtis}. 
Since $\Lambda^{\langle g\rangle}$ has rank $4$ and is primitive, 
$\Lambda^G=\Lambda^{\langle g\rangle}$ and the symmetry group $G$ must be the 
stabiliser $5^{1+2}:\ZZ_4$ of $\Lambda^{\langle g\rangle}$. 
\medskip

It was shown in \cite{Gaberdiel:2011fg} that the Gepner model $(1)^6$ 
corresponds to the case (ii) of the classification theorem. It thus follows from
the above reasoning that it must indeed 
be equivalent to a $\ZZ_3$-orbifold of $\mathbb{T}^4$, see also \cite{Fluder}. 
(We shall also study these orbifolds more systematically in section~\ref{sec:Z3}.)
In the next section, we will provide an explicit construction of a $\ZZ_5$-orbifold of a 
torus model and show that its symmetry group is $5^{1+2}:\ZZ_4$, as predicted
by the above analysis.

\section{A K3 model with symmetry group $5^{1+2}:\ZZ_4$}\label{sec:Z5}

In this section we will construct a supersymmetric  sigma-model on 
$\mathbb{T}^4$ with a symmetry $g$ of order $5$ commuting with an $\N=(4,4)$ 
superconformal algebra and acting asymmetrically on the left- and on the right-moving sector. 
The orbifold of this model by $g$ will turn out to be a well-defined SCFT with $\N=(4,4)$ (in 
particular, the level matching condition is satisfied) that can be interpreted as a non-linear 
sigma-model on K3. We will argue that the group of symmetries of this model is
$G=5^{1+2}.\ZZ_4$, one of the exceptional groups considered in the general classification 
theorem.

\subsection{The torus model}

Let us consider a supersymmetric  sigma-model on the torus 
$\mathbb{T}^4$. %=\RR^4/\Lambda$, where $\Lambda\in\RR^4$ is a lattice of rank $4$. 
Geometrically, we can characterise the theory in terms of a metric and 
a Kalb-Ramond field, but it is actually more convenient to describe it simply
as a conformal field theory that is generated by the following fields:
%for some lattice $\Lambda\in\RR^4$ of rank $4$, with a 
%Euclidean constant metric $\G$ and a constant antisymmetric background 
%Kalb-Ramond-field $B$. We 
%shall choose the coordinates on $\RR^4$ so that the metric is $\G_{ab}=\delta_{ab}$.
four left-moving $u(1)$ currents $\partial X^a(z)$, $a=1,\ldots,4$,  four free fermions 
$\psi^a(z)$, $a=1,\ldots,4$, their right-moving analogues 
$\bar\partial X^a(\bar z),\tilde\psi^a(\bar z)$, as well as some 
winding-momentum fields $V_\lambda(z,\bar z)$ that are associated to vectors 
$\lambda$ in an even
unimodular lattice $\Gamma^{4,4}$ of signature $(4,4)$. 
The mode expansions of the left-moving fields are
\be \partial X^a(z)=\sum_{n\in\ZZ} \alpha_n z^{-n-1}\ , \qquad 
\psi^a=\sum_{n\in\ZZ+\nu} \psi_n z^{-n-\frac{1}{2}}\ ,
\ee 
where $\nu=0,1/2$ in the R- and NS-sector, respectively. Furthermore, we have the usual 
commutation relations 
\be\label{commutations} 
{}[\alpha_m^a,\alpha_n^b]=m\, \delta^{ab}\, \delta_{m,-n}\qquad\qquad 
\{\psi_m^a,\psi_n^b\}=\delta^{ab}\, \delta_{m,-n}\ .
\ee 
Analogous statements also hold for 
the right-moving modes $\tilde\alpha_n$ and $\tilde \psi_n$. 
The vectors $\lambda\equiv (\lambda_L,\lambda_R)\in\Gamma^{4,4}$ describe the 
eigenvalues of the 
corresponding  state with  respect to the left- and right-moving zero modes 
$\alpha^a_0$ and  $\tilde\alpha^a_0$, respectively. In these conventions the inner
product on $\Gamma^{4,4}$ is given as 
\be
(\lambda,\lambda') = \lambda_L \cdot \lambda_L' - \lambda_R \cdot \lambda_R' \ .
\ee

\subsubsection{Continuous and discrete symmetries}

Any torus model contains an
$\hat{\mathfrak{su}}(2)_1\oplus \hat{\mathfrak{su}}(2)_1\oplus 
\hat{\mathfrak{u}}(1)^4$ current algebra, both on the left and on the right. Here, 
the $\hat{\mathfrak{u}}(1)^4$ currents are the $\partial X^a$, $a=1,\ldots,4$, while 
$\hat{\mathfrak{su}}(2)_1\oplus \hat{\mathfrak{su}}(2)_1=\hat{\mathfrak{so}}(4)_1$ 
is generated by the fermionic bilinears
\begin{align}\label{su(2)1} 
a^3:=\bar\psi^{(1)}\psi^{(1)}+\bar\psi^{(2)}\psi^{(2)}\qquad 
a^+:=\bar\psi^{(1)}\bar\psi^{(2)}\qquad a^-:=-\psi^{(1)}\psi^{(2)}\ ,\\
\label{su(2)2} 
\hat a^3:=\bar\psi^{(1)}\psi^{(1)}-\bar\psi^{(2)}\psi^{(2)}\qquad 
\hat a^+:=\bar\psi^{(1)}\psi^{(2)}\qquad \hat a^-:=-\psi^{(1)}\bar\psi^{(2)}\ ,
\end{align}
where
\begin{eqnarray}
\psi^{(1)}& =\tfrac{1}{\sqrt{2}}\, (\psi^1+i\psi^2)\qquad \qquad \qquad
& \psi^{(2)}=\tfrac{1}{\sqrt{2}}\, (\psi^3+i\psi^4)\\
\bar{\psi}^{(1)}& =\tfrac{1}{\sqrt{2}}\, (\psi^1-i\psi^2)\qquad \qquad \qquad
& \bar{\psi}^{(2)}=\tfrac{1}{\sqrt{2}}\, (\psi^3-i\psi^4)\ .
\end{eqnarray}
At special points in the moduli space, where the $\Gamma^{4,4}$ lattice contains vectors of the 
form $(\lambda_L,0)$ with $\lambda_L^2=2$, the bosonic $\mathfrak{u}(1)^4$ algebra is 
enhanced to some 
non-abelian algebra $\g$ of rank $4$. There are generically $16$ (left-moving) supercharges;
they form four $(\bf 2,\bf 2)$ representations of the 
${\mathfrak{su}}(2)\oplus {\mathfrak{su}}(2)$ zero mode algebra from 
(\ref{su(2)1}) and (\ref{su(2)2}). 
Altogether, the chiral algebra at 
generic points is a \emph{large} $\N=4$ superconformal algebra. 
\medskip

%These $16$ states form a spinor representation under the $SO(4,4,\RR)$ group that maps different torus models into one each other. In particular, they form two  $SO(4,4,\RR)$ Weyl spinors with opposite chiralities, where the chirality is the eigenvalue of
%\be (-1)^{F+\tilde F}:=\psi^1\cdots\psi^4\tilde\psi^1\dots\tilde\psi^4\ .
%\ee A R-R ground state  $|s_1,s_2;\tilde s_1,\tilde s_2\rangle$ is an eigenvector with eigenvalue $(-1)^{2(s_1+s_2+\tilde s_1+\tilde s_2)}$.
%Symmetries of the model correspond to dualities $O(4,4,\ZZ)$ (??? what about orientation reversing symmetries? to be understood ???) that are contained in the subgroup $SO(4)\times SO(4)\subset SO(4,4,\RR)$ that leave the spaces of left- and right-moving momenta set-wise invariant. In particular, $\psi_0^{(1)},\psi_0^{(2)},\bar\psi_0^{(1)},\bar\psi_0^{(2)},\tilde \psi_0^{(1)},\tilde\psi_0^{(2)},\tilde{\bar\psi_0^{(1)}},\tilde{\bar\psi_0^{(2)}},$ is a basis of eigenvectors with respect to $(J_3\pm \hat J_3)/2$ in \eqref{su21} and \eqref{su22} and their right-moving analogues, that form a Cartan algebra for $SO(4,4,\RR)$.

%\subsection{A symmetry of order $5$}

We want to construct a model with a symmetry $g$ of order $5$, acting non-trivially on the 
fermionic fields, and commuting with the \emph{small} $\N=4$ subalgebras both on the left and 
on the right. A small $\N=4$ algebra contains an $\hat{\mathfrak{su}}(2)_1$ current algebra and 
four supercharges in two doublets of $\mathfrak{su}(2)$. The symmetry $g$ acts by an 
${\rm O}(4,\RR)$ rotation on the left-moving fermions $\psi^a$, preserving the anti-commutation 
relations \eqref{commutations}. Without loss of generality, we may assume that $\psi^{(1)}$ and 
$\bar\psi^{(1)}$ are eigenvectors of $g$ with eigenvalues $\zeta$ and $\zeta^{-1}$, where 
$\zeta$ is a primitive fifth root of unity
\be \zeta^5=1\ ,
\ee
and that the $\hat{\mathfrak{su}}(2)_1$ algebra preserved by $g$ is \eqref{su(2)1}. 
This implies that $g$ acts on all the fermionic fields by
\be\label{fermtransf}  
\psi^{(1)}\mapsto \zeta\, \psi^{(1)}\ ,\quad  
\bar\psi^{(1)}\mapsto \zeta^{-1} \bar\psi^{(1)}\ ,\quad 
\psi^{(2)}\mapsto \zeta^{-1} \psi^{(2)}\ ,\quad 
\bar\psi^{(2)}\mapsto \zeta \, \bar\psi^{(2)}\ .
\ee  
Note that the action of $g$ on the fermionic fields can be described by
$e^{\frac{2\pi i k}{5}\hat a^3_0}$ for some 
$k=1,\ldots,4$, where $\hat a^3$ is the current in the algebra \eqref{su(2)2}. The four 
$g$-invariant supercharges can then be taken to be 
\be\label{bostransf} 
\sqrt{2}\sum_{i=1}^2J^{(i)}\bar \psi^{(i)}\ , \quad 
\sqrt{2}\sum_{i=1}^2\bar J^{(i)}\psi^{(i)}\ , \quad
\sqrt{2}(\bar J^{(1)}\bar \psi^{(2)}-\bar J^{(2)}\bar \psi^{(1)})\ , \quad 
\sqrt{2}(J^{(1)}\psi^{(2)}-J^{(2)}\psi^{(1)})\ ,
\ee 
where $J^{(1)},\bar J^{(1)},J^{(2)},\bar J^{(2)}$ are suitable (complex) linear 
combinations of the  left-moving currents $\partial X^a$, $a=1,\ldots,4$.
%form an orthonormal basis for the 
%complexification $\Pi_L\otimes\CC$ of the space of bosonic left-moving currents 
%$\partial X^a$, $a=1,\ldots,4$.
%
%\be \partial Z^{(1)}=2^{-1/2}(\partial X^1+i\partial X^2)\ ,\qquad\qquad J^{(2)}=2^{-1/2}(J^3+iJ^4)\ ,
%\ee and $J^1,\ldots,J^4$ generate a $u(1)^4$ subalgebra of the bosonic current algebra $\g$.
In order to preserve the four supercharges, $g$ must act with the same eigenvalues on 
the bosonic currents 
\be\label{bostransf2} J^{(1)}\mapsto \zeta \, J^{(1)}\ ,\quad  
\bar J^{(1)}\mapsto \zeta^{-1} \bar J^{(1)}\ ,\quad 
J^{(2)}\mapsto \zeta^{-1} J^{(2)}\ ,\quad 
\bar J^{(2)}\mapsto \zeta \, \bar J^{(2)}\ .
\ee 
A similar reasoning applies to the right-moving algebra with respect to an eigenvalue 
$\tilde\zeta$, with $\tilde\zeta^5=1$. For the symmetries with a geometric interpretation, 
the action on the left- and right-moving bosonic currents is induced by an 
${\rm O}(4,\RR)$-transformation on the scalar fields $X^a$, $a,=1,\ldots,4$, representing 
the coordinates on the torus; then $\zeta$ and $\tilde\zeta$ are necessarily equal. In our 
treatment, we want to allow for the more general case where $\zeta\neq \tilde\zeta$.
\smallskip

The action of $g$ on $J^a$ and $\tilde J^a$ induces an ${\rm O}(4,4,\RR)$-transformation 
on the lattice $\Gamma^{4,4}$. % of $(\alpha_0,\tilde\alpha_0)$ eigenvalues. 
The transformation $g$ is a symmetry of the model if and only if it induces an automorphism on 
$\Gamma^{4,4}$. In particular, it must act by an invertible integral matrix on any lattice basis. The requirement 
that the trace of this matrix (and of any power of it) must be integral, leads to the
condition that 
\be\label{zetaint}
2(\zeta^i+\zeta^{-i}+\tilde\zeta^i+\tilde \zeta^{-i})\in\ZZ\ ,
\ee 
for all $i\in\ZZ$. For $g$ of order $5$, this condition is satisfied by
\be\label{eigenv} 
\zeta=e^{\frac{2\pi i}{5}}\qquad\hbox{and} \qquad \tilde \zeta=e^{\frac{4\pi i}{5}}\ ,
\ee 
and this solution is essentially unique (up to taking powers of it or exchanging the 
definition of $\zeta$ and $\zeta^{-1}$). Eq.~\eqref{eigenv} shows that a supersymmetry
preserving symmetry of 
order $5$ is necessarily left-right asymmetric, and hence does not have a geometric 
interpretation. 
\medskip

It is now clear how to construct a torus model with the symmetries \eqref{fermtransf} and
\eqref{bostransf2}. First of all, we need an automorphism $g$ of $\Gamma^{4,4}$ of order five. 
Such an automorphism must have eigenvalues $\zeta,\zeta^2,\zeta^3,\zeta^4$, each 
corresponding to two independent eigenvectors $v_{\zeta^i}^{(1)},v_{\zeta^i}^{(2)}$, 
$i=1,\ldots, 4$, in $\Gamma^{4,4}\otimes\CC$. 
Given the discussion above, see in particular  \eqref{eigenv}, 
we now require that the vectors
\be
v_{\zeta^1}^{(1)}\ ,\quad  v_{\zeta^1}^{(2)}\ , \quad
v_{\zeta^4}^{(1)}\ , \quad v_{\zeta^4}^{(2)}
\ee
span a positive-definite subspace of $\Gamma^{4,4}\otimes\CC$ (i.e.\ correspond
to the left-movers), while the vectors
\be
v_{\zeta^2}^{(1)}\ ,\quad 
v_{\zeta^2}^{(2)}\ ,\quad
v_{\zeta^3}^{(1)}\ , \quad
v_{\zeta^3}^{(2)} 
\ee
span a negative-definite subspace of $\Gamma^{4,4}\otimes\CC$ (i.e.\ correspond to 
the right-movers). 
%is endowed with the sesquilinear form induced by 
%the bilinear form on $\Gamma^{4,4}$). If such an automorphism exists, there is a  torus model 
%where $\Pi_L\otimes\CC$ is spanned by $v_{\zeta^1}^{(1)}$, 
%$v_{\zeta^1}^{(2)}$, $v_{\zeta^4}^{(1)}$, $v_{\zeta^4}^{(2)}$, and $\Pi_R\otimes\CC$ by 
%$v_{\zeta^2}^{(1)}$ $v_{\zeta^2}^{(2)}$, $v_{\zeta^3}^{(1)}$, $v_{\zeta^3}^{(2)}$. Then, 
%the action of $g$ can be extended to a symmetry $g$ acting as in \eqref{fermtransf} and 
%\eqref{bostransf} on the bosonic and fermionic fields.
%\bigskip

An automorphism $g$ with the properties above can be explicitly constructed as follows. 
Let us consider the real vector space with basis vectors
$x_1,\ldots,x_4$, and $y_1,\ldots,y_4$, and define a linear map $g$ of order 
$5$ by 
\be g(x_i)=x_{i+1}\ ,\qquad g(y_i)=y_{i+1}\ ,\qquad\qquad i=1,\ldots,3\ ,
\ee and
\be g(x_4)=-(x_1+x_2+x_3+x_4)\ ,\qquad g(y_4)=-(y_1+y_2+y_3+y_4)\ .
\ee
%\be \lambda_1+g(\lambda_1)+g^2(\lambda_1)+g^3(\lambda_1)+g^4(\lambda_1)=0=\mu_1+g(\mu_1)+g^2(\mu_1)+g^3(\mu_1)+g^4(\mu_1)\ .
%\ee 
A $g$-invariant bilinear form on the space is uniquely determined by the conditions
\be 
(x_i,x_i)=0=(y_i,y_i)\ ,\qquad i=1,\ldots, 4
\ee 
and
\be 
(x_1, x_2)=1 \ ,\quad (x_1,x_3)=(x_1,x_4)=-1 \ , \qquad
 (y_1, y_2)=1\ , \quad (y_1,y_3)=(y_1,y_4)=-1 \ ,
\ee
as well as 
\be
(x_1,y_1)=1\ , \quad \hbox{and} \quad (x_i,y_1)=0\ , \ \ (i=2,3,4)\ .
\ee 
The lattice spanned by these basis vectors is an indefinite even unimodular lattice of rank $8$ 
and thus necessarily isomorphic to $\Gamma^{4,4}$. The $g$-eigenvectors can be easily 
constructed in terms of the  basis vectors
%\be v^{(1)}_{\zeta^i}:=\sum_{j=0}^4 \zeta^{-ij}g^j(x_1+y_1)\ ,\qquad v^{(2)}_{\zeta^i}:=\sum_{j=0}^4 \zeta^{-ij}g^j(x_1-y_1)\ ,
%\ee
 and one can verify that the eigenspaces have the 
correct signature. 

This torus model has an additional $\ZZ_4$ symmetry group that preserves the 
superconformal algebra and normalises the group generated by $g$. The generator $h$ 
of this group acts by
\begin{align}\label{haction3a} %g^i(x_1)\mapsto g^{i+3}(y_{1})\ ,\qquad g^{i}(y_1)\mapsto -g^{i+2}(x_{1})\ ,\qquad\qquad i=0,\ldots,4\ ,
& h(x_i):=g^{1-i}(x_1+x_4+2y_1+y_2+y_3+y_4)\ ,\\ \label{haction3b} &
h(y_i):=g^{1-i}(-2x_1-x_2-x_3-x_4-y_1-y_3-y_4)\ ,\qquad i=1,\ldots,4\ ,
\end{align}
on the lattice vectors.
The $g$-eigenvectors $v^{(a)}_{\zeta^i}$, $a=1,2$, $i=1,\ldots,4$ can be defined as
\be v^{(1)}_{\zeta^i}:=\sum_{j=0}^4 \zeta^{-ij}g^j(x_1+h(x_1))\ ,\qquad 
v^{(2)}_{\zeta^i}:=\sum_{j=0}^4 \zeta^{-ij}g^j(x_1-h(x_1))\ ,
\ee
so that 
\be 
h(v^{(1)}_{\zeta^i})=-v^{(2)}_{\zeta^{-i}}\ ,\qquad h(v^{(2)}_{\zeta^i})=v^{(1)}_{\zeta^{-i}}\ .
\ee 
Correspondingly, the action of $h$ on the left-moving fields is
\begin{align}\label{haction1} 
&\psi^{(1)}\mapsto -\psi^{(2)}\ ,\quad \psi^{(2)}\mapsto \psi^{(1)}\ ,\quad 
\bar\psi^{(1)}\mapsto -\bar\psi^{(2)}\ ,\quad \bar\psi^{(2)}\mapsto \bar\psi^{(1)}\ ,\\ \label{haction2}
&J^{(1)}\mapsto -J^{(2)}\ ,\quad J^{(2)}\mapsto J^{(1)}\ ,\quad \bar J^{(1)}\mapsto -\bar J^{(2)}\ ,
\quad \bar J^{(2)}\mapsto \bar J^{(1)}\ ,
\end{align}
 and the action on the right-moving fields is analogous. It is immediate to verify that the 
 generators of the superconformal algebra are invariant under this transformation. 
 %On the lattice vectors, $h$ acts by
%\be\label{haction3} g^i(x_1)\mapsto g^{-i}(y_{1})\ ,\qquad g^{i}(y_1)\mapsto -g^{-i}(x_{1})\ ,\qquad\qquad i=0,\ldots,4\ .
%\ee 

%one can choose the $g$-eigenvectors $v^{(a)}_{\zeta^i}$, $a=1,2$, $i=1,\ldots,4$, in such a way 
%that 
%\be 
%h(v^{(1)}_{\zeta^i})=v^{(2)}_{\zeta^{-i}}\ ,\qquad h(v^{(2)}_{\zeta^i})=-v^{(1)}_{\zeta^{-i}}\ .
%\ee 
%Correspondingly, the action of $h$ on the left-moving fields is
%\begin{align}\label{haction1} &\psi^{(1)}\mapsto \psi^{(2)}\ ,\quad \psi^{(2)}\mapsto -\psi^{(1)}\ ,
%\quad \bar\psi^{(1)}\mapsto \bar\psi^{(2)}\ ,\quad \bar\psi^{(2)}\mapsto -\bar\psi^{(1)}\ ,\\ 
%\label{haction2}
%&J^{(1)}\mapsto J^{(2)}\ ,\quad J^{(2)}\mapsto -J^{(1)}\ ,\quad \bar J^{(1)}\mapsto \bar J^{(2)}\ ,
%\quad \bar J^{(2)}\mapsto -\bar J^{(1)}\ ,
%\end{align}
% and the action on the right-moving fields is analogous. 
 %On the lattice vectors, $h$ acts by
%\be\label{haction3} g^i(x_1)\mapsto g^{-i}(y_{1})\ ,\qquad g^{i}(y_1)\mapsto -g^{-i}(x_{1})\ ,\qquad
%\qquad i=0,\ldots,4\ .
%\ee 

\subsection{The orbifold theory}

Next we want to consider the orbifold of this torus theory by the group $\mathbb{Z}_5$ 
that is generated by $g$.

\subsubsection{The elliptic genus}

The elliptic genus of the orbifold theory can be computed by summing over the 
${\rm SL}(2,\ZZ)$ images of the untwisted sector contribution, which in turn is given by
\be 
\phi^U(\tau,z)= \frac{1}{5}\sum_{k=0}^4\phi_{1,g^k}(\tau,z)\ ,
\ee
where
\be \phi_{1,g^k}(\tau,z)
=\hbox{Tr}_{\rm RR}(g^k \, q^{L_0-\frac{c}{24}}\bar q^{\tilde L_0-\frac{\tilde c}{24}}
y^{2J_0}(-1)^{F+\tilde F})\ .
\ee 
The $k=0$ contribution, i.e.\ the elliptic genus of the original torus theory, is zero. Each
$g^k$-contribution, for $k=1,\ldots,4$, is the product of a factor coming from the ground 
states, one from the oscillators and one from the momenta
\be 
\phi_{1,g^k}(\tau,z)=\phi_{1,g^k}^{\rm gd}(\tau,z)\,
\phi_{1,g^k}^{\rm osc}(\tau,z)\,
\phi_{1,g^k}^{\rm mom}(\tau,z)\ .
\ee 
These contributions are, respectively,
\be 
\phi_{1,g^k}^{\rm gd}(\tau,z)= y^{-1}(1-\zeta^k y)(1-\zeta^{-k} y)(1-\zeta^{2k})(1-\zeta^{-2k})
=2y^{-1}+2y+1\ ,
\ee
\be 
\phi_{1,g^k}^{\rm osc}(\tau,z)=
\prod_{n=1}^\infty \frac{(1-\zeta^k yq^n)(1-\zeta^{-k} yq^n)(1-\zeta^{k}y^{-1}q^n)
(1-\zeta^{-k}y^{-1}q^n)}{(1-\zeta^k q^n)^2(1-\zeta^{-k}q^n)^2}\ ,
\ee 
and
\be 
\phi_{1,g^k}^{\rm mom}(\tau,z)=1\ ,
\ee 
where the last result follows because the only $g$-invariant state of the form 
$(k_L,k_R)$ is the vacuum $(0,0)$. Thus we have 
\be 
\phi_{1,g^k}(\tau,z)=5\, \frac{\vartheta_1(\tau,z+\frac{k}{5})\,
\vartheta_1(\tau,z-\frac{k}{5})}{\vartheta_1(\tau,\frac{k}{5})\, \vartheta_1(\tau,-\frac{k}{5})}\ ,
\ee 
where 
\be 
\vartheta_1(\tau,z)=-iq^{1/8}y^{-\frac12}(y-1)\prod_{n=1}^\infty (1-q^n)(1-yq^n)(1-y^{-1}q^n)\ ,
\ee 
is the first Jacobi theta function. Modular transformations of $\phi_{1,g^k}(\tau,z)$ reproduce 
the twining genera in the twisted sector
\be \label{Z5twis}
\phi_{g^l,g^k}(\tau,z)=\hbox{Tr}_{\Hh^{(l)}} \bigl(g^k \, q^{L_0-\frac{c}{24}}\,
\bar q^{\tilde L_0-\frac{\tilde c}{24}}y^{2J_0}(-1)^{F+\tilde F}\bigr)\ ,
\ee
and using the modular properties of the theta function we obtain
\be 
\phi_{g^l,g^k}(\tau,z)=5\, \frac{\vartheta_1(\tau,z+\frac{k}{5}+\frac{l\tau}{5})\,
\vartheta_1(\tau,z-\frac{k}{5}-\frac{l\tau}{5})}{\vartheta_1(\tau,\frac{k}{5}+\frac{l\tau}{5})\,
\vartheta_1(\tau,-\frac{k}{5}-\frac{l\tau}{5})}\ ,
\ee 
for $k,l\in\ZZ/5\ZZ$, $(k,l)\neq (0,0)$.
The elliptic genus of the full orbifold theory is then
\be 
\phi_{\rm orb}(\tau,z)=\frac{1}{5}\sum_{k,l\in\ZZ/5\ZZ} \phi_{g^l,g^k}(\tau,z)
=\sum_{\substack{k,l\in\ZZ/5\ZZ\\(k,l)\neq (0,0)}}\frac{\vartheta_1(\tau,z+\frac{k}{5}+\frac{l\tau}{5})\,
\vartheta_1(\tau,z-\frac{k}{5}-\frac{l\tau}{5})}{\vartheta_1(\tau,\frac{k}{5}+\frac{l\tau}{5})\,
\vartheta_1(\tau,-\frac{k}{5}-\frac{l\tau}{5})}\ .
\ee 
Since $\phi_{g^k,g^l}(\tau,0)=5$ for all $(k,l)\neq (0,0)$, we have
\be \label{indexorb}
\phi_{\rm orb}(\tau,0)=\frac{1}{5}\sum_{\substack{k,l\in\ZZ/5\ZZ\\(k,l)\neq (0,0)}} 5 =24\ ,
\ee 
which shows that the orbifold theory is a non-linear sigma-model on K3. In particular, the 
untwisted sector has $4$ RR ground states, while each of the four twisted sectors contains 
$5$ RR ground states. For the following it will be important to understand the structure
of the various twisted sectors in detail.

\subsubsection{The twisted sectors}\label{sec:twistsect}

In the $g^k$-twisted sector, let us consider a basis of $g$-eigenvectors for the currents 
and fermionic fields. For a given eigenvalue $\zeta^{i}$, $i\in\ZZ/5\ZZ$, of $g^k$, the 
corresponding currents $J^{i,a}$ and fermionic fields $\psi^{i,b}$ (where $a,b$ labels 
distinct eigenvectors with the same eigenvalue) have a mode expansion
\be 
J^{i,a}(z)=\sum_{n\in \frac{i}{5}+\ZZ} \alpha^{i,a}_nz^{-n-1}\ ,\qquad 
\psi^{i,a}(z)=\sum_{r\in \frac{i}{5}+\nu+\ZZ}\psi^{i,a}_rz^{-r-1/2}\ ,
\ee 
where $\nu=1/2$ in the NS- and $\nu=0$ in the R-sector. The ground states of 
the $g^k$-twisted sector are characterised by the conditions
\begin{align} 
&\alpha^{i,a}_n|m,k\rangle=\tilde\alpha^{i,a}_n|m,k\rangle = 0 \ ,\qquad 
&\forall\ n>0,\ i,\ a\ ,\\ 
&\psi^{i,b}_r|m,k\rangle=\tilde\psi^{i,b}_r|m,k\rangle = 0 \ ,\qquad &\forall\ r>0,\ i,\ b\ ,
\end{align} 
where $|m,k\rangle$ denotes the $m^{\rm th}$ ground state in the $g^k$-twisted sector. 
Note that since none of the currents are $g$-invariant, there are no current zero modes
in the $g^k$-twisted sector, and similarly for the fermions. For a given $k$, the states 
$|m,k\rangle$ have then all the same conformal dimension, which can be calculated using 
the commutation relation $[L_{-1},L_1]=2L_0$ or read off from the leading term of the
modular transform of the twisted character (\ref{Z5twis}). In the $g^k$-twisted NS-NS-sector 
the ground states have conformal dimension
\be
\hbox{NS-NS $g^k$-twisted:}\qquad 
h=\frac{k}{5} \qquad \hbox{and} \qquad  \tilde h=\frac{2k}{5} \ , 
\ee
while in the RR-sector we have instead 
\be
\hbox{R-R $g^k$-twisted:}\qquad 
h=\tilde{h} = \frac{1}{4} \ . 
\ee
In particular, level matching is satisfied, and thus the asymmetric orbifold is consistent
\cite{Narain:1986qm}. The full $g^k$-twisted sector is then obtained by acting with 
the negative modes of the currents and the fermionic fields on the ground states
$|m,k\rangle$.
\smallskip

Let us have a closer look at the ground states of the $g^k$ twisted sector; for concreteness
we shall restrict ourselves to the case $k=1$, but the modifications for general $k$ are minor
(see below). The vertex operators $V_\lambda(z,\bar z)$ associated to 
$\lambda\in\Gamma^{4,4}$, act on the ground states $|m,1\rangle$ by
\be 
\lim_{z\to 0} V_\lambda(z,\bar z) |m,1\rangle =e_{\lambda} |m,1\rangle\ ,
\ee 
where $e_\lambda$ are operators commuting with all current and fermionic oscillators 
and satisfying
\be\label{ecomm} 
e_\lambda \, e_\mu = \epsilon(\lambda,\mu)\, e_{\lambda+\mu}\ ,
\ee 
for some fifth root of unity $\epsilon(\lambda,\mu)$. The vertex operators
$V_\lambda$ and $V_\mu$ must be local relative to one another, and this is the case
provided that (see the appendix) 
\be  
\frac{\epsilon(\lambda,\mu)}{\epsilon(\mu,\lambda)} = C(\lambda,\mu) \ ,
\ee  
where
\be \label{Pgdef} 
C(\lambda,\mu)=%(-1)^{\sum_{i=0}^4 g^i(\lambda)\cdot\mu}
\prod_{i=1}^4(\zeta^i)^{(g^i(\lambda),\mu)}=\zeta^{(P_g(\lambda),\mu)}\qquad
\hbox{with} \qquad 
P_g(\lambda) = \sum_{i=1}^4 ig^i(\lambda)\ .
\ee 
The factor $C(\lambda,\mu)$ has the properties
\begin{align}\label{Clinear} 
&C(\lambda,\mu_1+\mu_2)=C(\lambda,\mu_1)\, C(\lambda,\mu_2)\ ,\qquad 
C(\lambda_1+\lambda_2,\mu)=C(\lambda_1,\mu)\, C(\lambda_2,\mu)\ ,\\  
&C(\lambda,\mu)=C(\mu,\lambda)^{-1}\ ,\label{Cinverse}\\ 
&C(\lambda,\mu)=C(g(\lambda),g(\mu))\ .\label{Cinvar}
\end{align}
Because of \eqref{Clinear}, $C(\lambda,0)=C(0,\lambda)=1$ for all 
$\lambda\in\Gamma^{4,4}$, and we can set
\be 
e_0=1\ ,
\ee 
so that $\epsilon(0,\lambda)=1=\epsilon(\lambda,0)$.
More generally, for the vectors $\lambda$ in the sublattice
\be 
R:=\{\lambda\in\Gamma^{4,4}\mid C(\lambda,\mu)=1\ ,\text{ for all } 
\mu\in\Gamma^{4,4}\}\subset\Gamma^{4,4}\ ,
\ee
we have $C(\lambda+\mu_1,\mu_2)=C(\mu_1,\mu_2)$, for all $\mu_1,\mu_2\in\Gamma^{4,4}$,
so that we can set
\be\label{Rinvariance} e_{\mu+\lambda}=e_\mu \ ,\qquad \forall \lambda\in R\ ,
\quad \mu\in\Gamma^{4,4}\ .
\ee 
Thus, we only need to describe the operators corresponding to representatives of the 
group $\Gamma^{4,4}/R$. The vectors $\lambda\in R$ are characterised by
\be 
(P_g(\lambda), \mu) \equiv 0\mod 5\ ,\qquad \text{ for all }\mu\in \Gamma^{4,4}\ ,
\ee
and since $\Gamma^{4,4}$ is self-dual this condition is equivalent to
\be 
P_g(\lambda)\in 5 \, \Gamma^{4,4}\ .
\ee 
Since $g$ has no invariant subspace, we have the identity
\be 
1+g+g^2+g^3+g^4=0
\ee 
that implies (see \eqref{Pgdef}) 
\be\label{Pgoneminusg} 
P_g\circ (1-g) = (1-g)\circ P_g =-5 \cdot \mathbf{1}\ .
\ee 
Thus, $\lambda\in R$ if and only if
\be 
P_g(\lambda)=P_g\circ(1-g)( \tilde\lambda)\ ,
\ee 
for some $\tilde\lambda\in\Gamma^{4,4}$,  and since $P_g$ has trivial kernel (see \eqref{Pgoneminusg}), we finally obtain
\be 
R=(1-g)\, \Gamma^{4,4}\ .
\ee
Since also $(1-g)$ has trivial kernel, $R$ has rank $8$ and $\Gamma^{4,4}/R$ is a finite 
group. Furthermore, 
\be 
|\Gamma^{4,4}/R|=\det(1-g)=25\ ,
\ee 
and, since $5\, \Gamma^{4,4}\subset R$, the group $\Gamma^{4,4}/R$ has exponent $5$. 
The only possibility is
\be 
\Gamma^{4,4}/R\cong \ZZ_5\times\ZZ_5\ .
\ee 
Let $x,y\in\Gamma^{4,4}$ be representatives for the generators of $\Gamma^{4,4}/R$. 
By \eqref{Cinverse}, we know that 
$C(x,x)=C(y,y)=1$, so that $C(x,y)\neq 1$ (otherwise $C$ would be 
trivial over the whole $\Gamma^{4,4}$), and we can choose $x,y$ such  that
\be 
C(x,y)=\zeta\ .
\ee  
Thus, the ground states form a representation of the algebra of operators generated by 
$e_x$, $e_y$, satisfying
\be 
e_x^5=1=e_y^5\ ,\qquad e_x e_y=\zeta e_y e_x\ .
\ee 
The group generated by $e_x$ and $e_y$ is the extra-special group $5^{1+2}$, and all 
its non-abelian irreducible representations\footnote{We call a representation non-abelian
if the central element does not act trivially.} 
are five dimensional.%The group of outer automorphism of $5^{1+2}$ is $\ZZ_5^\times\cong\ZZ_4$, where $k\in\ZZ_5^\times$ maps the central element $\zeta$ to $\zeta^k$. There are four non-abelian irreducible representations of $5^{1+2}$, distinguished by the character of the central element: they are all $5$-dimensional and related to one each other by outer automorphisms. 

In particular, for the representation on the $g$-twisted ground states, we can choose a basis of 
$e_x$-eigenvectors
\be 
|m;1\rangle \quad \hbox{with}  \quad 
e_x|m;1\rangle=\zeta^m|m;1\rangle\ , \qquad m\in \ZZ/5\ZZ\ ,
\ee  
and define the action of the operators $e_y$ by
\be 
e_y|m;1\rangle=|m+1;1\rangle\ .
\ee 
For any vector $\lambda\in\Gamma^{4,4}$, there are unique $a,b\in\ZZ/5\ZZ$ such that 
$\lambda=ax+by+(1-g)(\mu)$ for some  $\mu\in \Gamma^{4,4}$ and we 
define\footnote{The ordering of $e_x$ and $e_y$ in this definition is arbitrary; however, 
any other choice corresponds to a redefinition $\tilde e_\lambda=c(\lambda)e_\lambda$, 
for some fifth root of unity $c(\lambda)$, that does not affect the commutation relations 
$\tilde e_\lambda\tilde e_\mu=C(\lambda,\mu)\tilde e_\mu\tilde e_\lambda$.}
\be 
e_\lambda:=e_x^ae_y^b\ . 
\ee 
Since $g(\lambda)=ax+by+(1-g)(g(\mu)-ax-by)$, by \eqref{Rinvariance} we have
\be 
e_{g(\lambda)}=e_\lambda\ ,
\ee 
so that, with respect to the natural action $g(e_\lambda):=e_{g(\lambda)}$, the algebra is 
$g$-invariant. This is compatible with the fact that all ground states have the same left and 
right conformal weights $h$ and $\tilde h$, so that  the action of $g=e^{2\pi i(h-\tilde h)}$ 
is proportional to the identity.
\bigskip

The construction of the $g^k$-twisted sector, for $k=2,3,4$, is completely analogous to the 
$g^1$-twisted case, the only difference being that the root $\zeta$ in the definition of 
$C(\lambda,\mu)$ should be replaced by $\zeta^k$. Thus, one can define operators 
$e_x^{(k)},e_y^{(k)}$ on the $g^k$-twisted sector, for each $k=1,\ldots,4$, satisfying
\be\label{genrel}
(e_x^{(k)})^5=1=(e_y^{(k)})^5\ ,\qquad e_x^{(k)} e_y^{(k)}=\zeta^k e_y^{(k)} e_x^{(k)}\ .
\ee
The action of these operators on the analogous basis $|m;k\rangle$
with $m\in\ZZ/5\ZZ$ is then
\be \label{gkground} 
e_x^{(k)}|m;k\rangle=\zeta^{m}|m;k\rangle\ ,\quad e_y^{(k)}|m;k\rangle=|m+k;k\rangle\ .
\ee

\subsubsection{Spectrum and symmetries}\label{sec:twistsec2}

The spectrum of the actual orbifold theory is finally obtained from the above twisted
sectors by projecting onto the $g$-invariant states; technically, this is equivalent to
restricting to the states for which the difference of the left- and 
right- conformal dimensions is integer, $h-\tilde h\in\ZZ$. In particular, the 
RR ground states \eqref{gkground} in each (twisted) sector have $h=\tilde h=1/4$, so that 
they all
survive the projection. Thus, the orbifold theory has $4$ RR ground states in the untwisted sector 
(the spectral flow generators), forming a $({\bf 2},{\bf 2})$ representation of 
$\mathfrak{su}(2)_L \oplus \mathfrak{su}(2)_R$, and $5$ RR ground states in each twisted 
sector, which are singlets of $\mathfrak{su}(2)_L \oplus \mathfrak{su}(2)_R$.  In total
there are therefore $24$ RR ground states, as expected for a non-linear sigma-model 
on K3.  (Obviously, we are here just reproducing what we already saw in 
(\ref{indexorb}).)
\medskip

Next we want to define symmetry operators acting on the orbifold theory. First we can
construct operators $e_\lambda$ associated to $\lambda\in\Gamma^{4,4}$, that will form the 
extra special group $5^{1+2}$. They are defined to act by 
$e_\lambda^{(k)}$ on the $g^k$-twisted sector. The action of the untwisted sector preserves the subspaces $\Hh_m^{U}$, $m\in \ZZ/5\ZZ$, of states with momentum of the form $\lambda=nx+my+(1-g)(\mu)$,  for some $n\in\ZZ$ and $\mu\in \Gamma^{4,4}$.
% In the untwisted sector, these operators stabilise the subspaces $\Hh_m^{U}$, $m\in \ZZ/5\ZZ$ of states with momentum of the form $\lambda=nx+my+(1-g)(\mu)$,  for some $n\in\ZZ$ and $\mu\in \Gamma^{4,4}$.
%
%$e_y$ acts trivially, while $e_x$ multiplies a state with momentum $\lambda=nx+my+(1-g)(\mu)$,  for some $n\in\ZZ$ and $\mu\in \Gamma^{4,4}$, by a phase $\zeta^{m}$. %and trivially on the untwisted sector.
%In the untwisted sector, $e_x$ and $e_y$ commute and can be diagonalised simultaneously; the eigenstates with $e_x$-eigenvalue $\zeta^m$ and $e_y$-eigenvalue $\zeta^{-n}$ are the states with momentum of the form $\lambda=nx+my+(1-g)(\mu)$,   
%These operators enter in the definition of the untwisted vertex operators $\sum_i V_{g^i(\lambda)}(z,\bar z)$.
Let us denote by
$T_{m;k}$ a generic vertex operator associated with a $g^k$-twisted state, 
$k=1,\ldots,4$, with $e_x$-eigenvalue $\zeta^{m}$, $m\in\ZZ/5\ZZ$, and by
$T_{m;0}$ a vertex operator associated with a state in $\Hh_m^{U}$.   
Consistency 
of the OPE implies the 
fusion rules 
\be  \label{orbifusion}
T_{m;k}\times T_{m';k'}\to T_{m+m';k+k'}\ .
\ee  
These rules are preserved by the maps
\be 
T_{m;k}\mapsto e_\lambda \, T_{m;k} \, e_\lambda^{-1}\ ,\qquad
\lambda\in\Gamma^{4,4}\ ,
\ee 
which therefore define symmetries of the orbifold theory. As we have explained above,
these symmetries form the extra-special group $5^{1+2}$.

Finally, the symmetries \eqref{haction3a}, \eqref{haction3b}, \eqref{haction1} and \eqref{haction2}  
of the original torus theory induce a $\ZZ_4$-group of symmetries of the orbifold. Since 
$h^{-1}gh=g^{-1}$, the space of $g$-invariant states of the original torus theory is stabilised by 
$h$, so that $h$ restricts to a well-defined transformation on the untwisted sector of the orbifold. 
Furthermore, $h$ maps the $g^k$- to the $g^{5-k}$-twisted sector. Eqs.~\eqref{haction3a} and 
\eqref{haction3b} can be written as
\begin{align} &h(x_1)=2x_1+(1-g)(-x_1-x_2-x_3+y_1+y_2+y_3+y_4)\ ,\\ 
&h(y_1)=2y_1+(1-g)(-x_1-x_2-x_3-x_4-2y_1-y_2-y_3-y_4)\ .
\end{align} 
It follows that the action of  $h$ on the operators $e_\lambda^{(k)}$, $k=1,\ldots,4$ must be 
\be he_x^{(k)}h^{-1}=e_{2x}^{(5-k)}\ ,\qquad he_y^{(k)}h^{-1}=e_{2y}^{(5-k)}\ ,
\ee and it is easy to verify that this transformation is compatible with \eqref{genrel}.
 Correspondingly, the action on the twisted sector ground states is
\be h|m;k\rangle = |3m;5-k\rangle\ ,
\ee and it is consistent with \eqref{orbifusion}.

%can be associated to elements in the 
%(multiplicative) group 
%\be 
%\ZZ_5^\times:=\{\ell\in\ZZ/5\ZZ\text{ s.t. }\ell\neq 0\mod 5\}\ .
%\ee
%Indeed, for $\ell\in\ZZ_5^\times$ we can define the map
%\be 
%\ell: T_{m;k}\mapsto T_{m;\ell k}\ ,
%\ee  
%and this respects also the fusion rules (\ref{orbifusion}). 
Thus the full symmetry group is the
semi-direct product 
\be 
G=5^{1+2}:\ZZ_4\ ,
\ee 
where the generator $h\in\ZZ_4$ maps the central element 
$\zeta\in 5^{1+2}$ to $\zeta^{-1}$.

All of these symmetries act trivially on the superconformal algebra 
and on the spectral flow generators,
%(since they act trivially in the untwisted sector),
and therefore define symmetries in the sense of the  general classification theorem.
Indeed, $G$ agrees precisely 
with the group in case  (ii) of the theorem. Thus our orbifold theory 
realises this possibility.

\section{Models with symmetry group containing $3^{1+4}:\ZZ_2$}\label{sec:Z3}

Most of the torus orbifold construction described in the previous section generalises 
to symmetries $g$ of order different than $5$. In particular, one can show explicitly that  
orbifolds of $\TT^4$ models by a symmetry 
$g$ of order $3$ contain a group of symmetries $3^{1+4}:\ZZ_2$, so that they belong to 
one of the cases (iii) and (iv) of the theorem, as expected from the discussion in 
section~\ref{sec:symme}.

We take the action of the symmetry $g$ on the left-moving currents and fermionic fields 
to be of the form  \eqref{fermtransf} and \eqref{bostransf2}, where $\zeta$ is a now a 
third root of unity; analogous transformations hold for the right-moving fields with respect 
to a third root of unity $\tilde\zeta$. In this case, eq.\eqref{zetaint} can be satisifed by
\be 
\zeta=\tilde\zeta=e^{\frac{2\pi i}{3}}\ ,
\ee 
so that the action is left-right symmetric and $g$ admits an interpretation as a geometric 
${\rm O}(4,\RR)$-rotation of order $3$ of the torus $\TT^4$. For example, the torus 
$\RR^4/(A_2\oplus A_2)$, where $A_2$ is the root lattice of the $su(3)$ Lie algebra, 
with vanishing Kalb-Ramond field, admits such an automorphism. %Let $V\cong\RR^4$ 
%denote the vector representation of the group ${\rm O}(4,\RR)$ of rotations of the scalar 
%fields $X^a$. The space of marginal deformations preserving an $\N=(4,4)$ algebra 
%transforms in the $V\otimes V$ representation of ${\rm O}(4,\RR)$. In a model with 
%symmetry $g$ of order $3$ as above, there is an $8$-dimensional subspace of marginal 
%deformations which is fixed by $g$. It follows that in the moduli space of torus models, 
%there must be an $8$-dimensional family of models with a symmetry $g$ of order $3$, 
%whose tangent space at each point is spanned by the $g$-fixed marginal deformations.

\bigskip

The orbifold by $g$ contains $6$ RR ground states in the untwisted sector. In the 
$k^{\rm th}$ twisted sector, $k=1,2$, the ground states form a representation of an 
algebra of operators $e^{(k)}_{\lambda}$, $\lambda\in\Gamma^{4,4}$, satisfying 
the commutation relations
\be 
e^{(k)}_\lambda e^{(k)}_\mu=C(\lambda,\mu)^k e^{(k)}_\mu e^{(k)}_\lambda\ ,
\ee 
where 
\be 
C(\lambda,\mu) =\zeta^{(P_g(\lambda),\mu)}\ ,\qquad P_g=g+2g^2\ .
\ee 
As discussed in section~\ref{sec:twistsect}, we can set
\be 
e^{(k)}_{\lambda+\mu}=e_\mu^{(k)}\ , \qquad \forall \lambda\in R,\mu\in\Gamma^{4,4}\ ,
\ee 
where
\be 
R=(1-g)\Gamma^{4,4}\ .
\ee
(Note that $\Gamma^{4,4}$ contains no $g$-invariant vectors). The main difference with the 
analysis of section~\ref{sec:twistsect} is that, in this case,
\be 
\Gamma^{4,4}/R\cong\ZZ_3^4\ .
\ee 
In particular, we can find vectors $x_1,x_2,y_1,y_2\in\Gamma^{4,4}$ such that
\be 
C(x_i,y_j)=\zeta^{\delta_{ij}}\ ,\qquad C(x_i,x_j)=C(y_i,y_j)=1\ .
\ee 
The corresponding operators obey the relations
\be 
e^{(k)}_{x_i}e^{(k)}_{y_j}=\zeta^{k\delta_{ij}}e^{(k)}_{y_j}e^{(k)}_{x_i}\ ,\quad 
e^{(k)}_{x_i}e^{(k)}_{x_j}=e^{(k)}_{x_j}e^{(k)}_{x_i}\ ,\quad
e^{(k)}_{y_i}e^{(k)}_{y_j}=e^{(k)}_{y_j}e^{(k)}_{y_i}\ ,
\ee 
as well as
\be 
(e^{(k)}_{x_i})^3=1=(e^{(k)}_{y_i})^3\ .
\ee 
These operators generate the extra-special group $3^{1+4}$ of exponent $3$, and the 
$k^{\rm th}$-twisted ground states form a representation of this group. 
We can choose a basis $|m_1,m_2;k\rangle$, with $m_1,m_2\in \ZZ/3\ZZ$, 
of  simultaneous eigenvectors of $e^{(k)}_{x_1}$ and $e^{(k)}_{x_2}$, so that
\be 
e_{x_i}^{(k)}|m_1,m_2;k\rangle=\zeta^{m_i}|m_1,m_2;k\rangle\ ,\qquad 
e_{y_i}^{(k)}|m_1,m_2;k\rangle=|m_1+k\delta_{1i},m_2+k\delta_{2i};k\rangle\ .
\ee 
The resulting orbifold model has $9$ RR ground states in each twisted sector, for a total of $6+9+9=24$ RR ground states, as expected for a K3 model. As in section~\ref{sec:twistsec2}, the 
group $3^{1+4}$ generated by $e^{(k)}_\lambda$ extends to a group of symmetries of the 
whole orbifold model. Furthermore, the $\ZZ_2$-symmetry that flips the signs of the 
coordinates in the original torus theory induces a symmetry $h$ of the orbifold theory, 
which acts on the twisted sectors by
\be 
h\, |m_1,m_2;k\rangle = |-m_1,-m_2;k\rangle\ .
\ee
We conclude that the group $G$ of symmetries of any torus orbifold 
$\mathbb{T}^4/\ZZ_3$ contains a subgroup $3^{1+4}: \ZZ_2$, and is therefore included in the 
cases (iii) or (iv) of the classification theorem. This obviously  ties in nicely with the general
discussion of section~\ref{sec:symme}.

%As discussed in section~3.1,  the group 
%$3^{1+4}:\ZZ_2$ is a subgroup of $Co_0$ that stabilises an ${\cal S}$-lattice 
%$E_6^*(3)\subset\Lambda$. Each torus orbifold $\mathbb{T}^4/\ZZ_3$, corresponding to a 
%positive-definite subspace $\Pi\subset\Gamma^{4,20}$, can be associated with a 
%four-dimensional subspace $N\subset E_6^*(3)\otimes\RR\subset\Lambda\otimes\RR$, such that the lattice 
%$N^\perp\cap\Lambda$ is isomorphic (up to a sign flip in the quadratic form) to 
%$\Pi^\perp\cap \Gamma^{4,20}$. The Grassmannian parametrising such subspaces 
%$N\subset E_6^*(3)\otimes\RR$ is $8$-dimensional, so that it is reasonable to expect 
%this correspondence to be surjective. If this is the case, then all possible groups 
%considered in the cases (ii) and (iv) of the classification theorem can be realised 
%by torus orbifolds $\mathbb{T}^4/\ZZ_3$. The same  conclusion can be reached  within the 
%more rigorous approach of \cite{GV}, and is consistent with the  interpretation of the 
%Gepner model $(1)^6$, which realises the case (ii) of the classification theorem 
%\cite{Gaberdiel:2011fg} as a $\mathbb{T}^4/\ZZ_3$ orbifold.

\section{Conclusions}

In this paper we have reviewed the current status of the EOT conjecture
concerning a possible  $\mathbb{M}_{24}$ symmetry appearing in the elliptic genus of K3. 
We have explained that, in some sense, the EOT conjecture has already been proven since
twining genera, satisfying the appropriate modular and integrality properties, 
have been constructed for all conjugacy classes of $\mathbb{M}_{24}$.
However, the analogue of the Monster conformal field theory that would `explain' the 
underlying symmetry has not yet been found. In fact, no single K3 sigma-model
will be able to achieve this since none of them possesses
an automorphism group that contains $\mathbb{M}_{24}$. 

Actually, the situation is yet further complicated by the fact that there 
are K3 sigma-models whose automorphism group is {\em not even a 
subgroup of} $\ \mathbb{M}_{24}$; on the other hand, the elliptic genus of K3 
does not show any signs of exhibiting `Moonshine' with respect to any larger group.
As we have explained in this paper, most of the exceptional automorphism groups (i.e.\ 
automorphism groups that are not subgroups of $\mathbb{M}_{24}$) appear for K3s 
that are torus orbifolds. In fact, all cyclic torus orbifolds are necessarily exceptional in this
sense, and (cyclic) torus orbifolds account for all incarnations of the cases (ii) -- (iv) of the
classification theorem of \cite{Gaberdiel:2011fg} (see section~\ref{sec:symme}). We have checked 
these predictions by explicitly
constructing an asymmetric $\mathbb{Z}_5$ orbifold that realises case (ii) of 
the theorem (see section~\ref{sec:Z5}), and a family of $\mathbb{Z}_3$ orbifolds
realising cases (iii) and (iv) of the theorem (see section~\ref{sec:Z3}). Incidentally, 
these constructions also demonstrate that the exceptional cases (ii)-(iv) actually
appear in the K3 moduli space --- in the analysis of \cite{Gaberdiel:2011fg} this conclusion
relied on some assumption about the regularity of K3 sigma-models.
\smallskip

The main open problem that remains to be understood is why precisely
$\mathbb{M}_{24}$ is `visible' in the elliptic genus of K3, rather than any smaller
(or indeed bigger) group. Recently, we have constructed (some of) the twisted twining 
elliptic genera of K3 \cite{GPRV}, i.e.\ the analogues of Simon Norton's generalised 
Moonshine functions \cite{N}. We hope that they will help to shed further light on this 
question. 

\bigskip

\section*{Acknowledgements}
We thank Terry Gannon, Gerald Hoehn, and Michael Tuite for useful conversations. 
The research of MRG is supported in parts by the Swiss National Science
Foundation.

\appendix

\section{Commutation relations in the twisted sector}

The vertex operators $V_\lambda(z,\bar z)$ in the $g$-twisted sector can be defined in terms of formal exponentials of current oscillators%\footnote{We use the notation \be
%\sum_{\pm r>0} \lambda_L\cdot\alpha_r=\sum_{i=1}^4\sum_{\substack{r>0\\r\in\ZZ+\frac{i}{5}}}\sum_a\lambda_L^{i,a} \alpha_r^{i,a}\ ,\ee where $\alpha^{i,1},\alpha^{i,2},\ldots$ is a basis for the eigenspace of $g$ relative to the eigenvalue $\zeta^i$.}
\be E^{\pm}_\lambda(z,\bar z):= \exp\Bigl(\sum_{\substack{r\in \frac15 \ZZ\\ \pm r>0}}(\lambda_L\cdot \alpha)^{(r)}_{r}\frac{z^{-r}}{r}\Bigr)\exp\Bigl(\sum_{\substack{r\in\frac15\ZZ\\ \pm r>0}}(\lambda_R\cdot \tilde\alpha)^{(r)}_{r}\frac{\bar z^{-r}}{r}\Bigr)\ ,
%
%
%\exp\Bigl(\sum_{\pm r>0} \lambda_L\cdot\alpha_r\frac{z^{-r}}{r}\Bigr)\exp\Bigl(\sum_{\pm r>0} \lambda_R\cdot \tilde \alpha_r\frac{{\bar z}^{-r}}{r}\Bigr)\ .
\ee  where $(\lambda_L\cdot \alpha)^{(r)}_{r}$ and $(\lambda_R\cdot \tilde\alpha)^{(r)}_{r}$ are the $r$-modes of the currents
\begin{align} (\lambda_L\cdot \partial X)^{(r)}&:=\frac15\sum_{i=0}^4\zeta^{5ir}\lambda_L\cdot g^i(\partial X)=\frac15\sum_{i=0}^4\zeta^{5ir}g^{-i}(\lambda_L)\cdot \partial X\ ,\\
(\lambda_R\cdot \bar \partial X)^{(r)}&:=\frac15\sum_{i=0}^4\bar\zeta^{5ir}\lambda_R\cdot g^i(\bar\partial X)=\frac15\sum_{i=0}^4\bar \zeta^{5ir}g^{-i}(\lambda_R)\cdot \bar\partial X\ .
\end{align}
Then we can define
\be 
V_\lambda(z,\bar z):=E^-_\lambda(z,\bar z) \, E^+_\lambda(z,\bar z)\, e_\lambda\ ,
\ee 
where the operators $e_\lambda$ commute with all current oscillators and satisfy
\be 
e_\lambda \, e_\mu =\epsilon(\lambda,\mu) \, e_{\lambda+\mu}\ ,
\ee 
for some fifth root of unity $\epsilon(\lambda,\mu)$. The commutator factor
\be 
C(\lambda,\mu):=\frac{\epsilon(\lambda,\mu)}{\epsilon(\mu,\lambda)}\ ,
\ee 
can be determined by imposing the locality condition
\be 
V_\lambda(z_1,\bar z_1)\, V_\mu(z_2,\bar z_2)
=V_\mu(z_2,\bar z_2)\, V_\lambda(z_1,\bar z_1)\ .
\ee 
To do so, we note that the commutation relations between the operators 
$E^\pm_\lambda$ can be computed, as in \cite{Lepowsky1985},
 using the Campbell-Baker-Hausdorff formula 
\begin{multline}\label{commuta} 
E^{+}_\lambda(z_1,\bar z_1)E^{-}_\mu(z_2,\bar z_2)
= E^{-}_\mu(z_2,\bar z_2)E^{+}_\lambda(z_1,\bar z_1)\\ 
\prod_{i=0}^4 [(1-\zeta^{-i}(\tfrac{z_1}{z_2})^{\frac15})^{g^i(\lambda)_L\cdot\mu_L}
(1-\bar\zeta^{-i}(\tfrac{\bar z_1}{\bar z_2})^{\frac15})^{g^i(\lambda)_R\cdot\mu_R}]\ .
\end{multline} 
Using \eqref{commuta} and $e_\lambda e_\mu=C(\lambda,\mu)e_\mu e_\lambda$, the 
locality condition is then equivalent to
\be 
C(\lambda,\mu)\prod_{i=0}^4\frac{(1-\zeta^{-i}(\frac{z_1}{z_2})^{\frac15})^{g^i(\lambda)_L\cdot\mu_L}(1-\bar\zeta^{-i}\frac{\bar z_1^{1/5}}{\bar z_2^{1/5}})^{g^i(\lambda)_R\cdot\mu_R}}{(1-\zeta^{i}(\tfrac{z_2}{z_1})^{\frac15})^{g^i(\lambda)_L\cdot\mu_L}(1-\bar\zeta^{i}(\tfrac{\bar z_2}{\bar z_1})^{\frac15})^{g^i(\lambda)_R\cdot\mu_R}}=1\ ,
\ee 
that is
\be 
C(\lambda,\mu)\Bigl(-\frac{z_1^{1/5}}{z_2^{1/5}}
\Bigr)^{\sum_i g^i(\lambda)_L\cdot\mu_L}\Bigl(-\frac{\bar z_1^{1/5}}{\bar z_2^{1/5}}\Bigr)^{\sum_i g^i(\lambda)_R\cdot\mu_R}\prod_{i=0}^4[(\zeta^{-i})^{g^i(\lambda)_L\cdot\mu_L}(\bar\zeta^{-i})^{g^i(\lambda)_R\cdot\mu_R}]=1\ .
\ee 
Since $\Gamma^{4,4}$ has no $g$-invariant vector, we have the identities 
\be
\sum_{i=0}^4g^i(\lambda)_L=0=\sum_{i=0}^4g^i(\lambda)_R\ ,
\ee 
and hence finally obtain
\be 
C(\lambda,\mu)=\prod_{i=0}^4(\zeta^i)^{g^i(\lambda_L)\cdot\mu_L-g^i(\lambda_R)\cdot\mu_R}\ .
\ee

\end{document}